\documentclass[prd,twocolumn]{revtex4}

\pdfoutput=1

\usepackage{graphicx}
\usepackage{dcolumn}
\usepackage{bm}
\usepackage{amssymb,amsmath,bm}  
\usepackage{multirow}
\usepackage[utf8]{inputenc}
\usepackage{balance}
\usepackage{enumitem}

\usepackage{array}
\newcolumntype{C}[1]{>{\centering\let\newline\\\arraybackslash\hspace{0pt}}m{#1}}

\usepackage[usenames,dvipsnames]{xcolor}
\usepackage{hyperref}
\hypersetup{
    colorlinks = true,
    citecolor = {MidnightBlue},
    linkcolor = {BrickRed},
    urlcolor = {BrickRed}
}

\newcommand{\uKam}{\mu\text{K-arcmin}}
\newcommand{\nv}{\hat{\bf n}}

\begin{document}
\title{Reconstructing cosmic growth with kSZ observations \\in the era of Stage IV experiments}
\author{David Alonso$^1$, Thibaut Louis$^2$, Philip Bull$^{3,4}$, Pedro G. Ferreira$^1$}
\affiliation{$^{1}$University of Oxford, Denys Wilkinson Building,
             Keble Road, Oxford, OX1 3RH,  UK\\
             $^{2}$UPMC Univ Paris 06, UMR7095, Institut d'Astrophysique de Paris, F-75014,
                    Paris, France\\
             $^{3}$California Institute of Technology, Pasadena, CA 91125, USA\\
             $^{4}$Jet Propulsion Laboratory, California Institute of Technology,
                   4800 Oak Grove Drive, Pasadena, California, USA}

\begin{abstract}
  Future ground-based CMB experiments will generate competitive large-scale structure datasets by
  precisely characterizing CMB secondary anisotropies over a large fraction of the sky. We describe
  a method for constraining the growth rate of structure to sub-1\% precision out to $z\approx 1$,
  using a combination of galaxy cluster peculiar velocities measured using the kinetic
  Sunyaev-Zel'dovich (kSZ) effect, and the velocity field reconstructed from galaxy redshift
  surveys. We consider only thermal SZ-selected cluster samples, which will consist of
  $\mathcal{O}(10^4-10^5)$ sources for Stage 3 and 4 CMB experiments respectively. Three
  different methods for separating the kSZ effect from the primary CMB are compared, including a
  novel blind ``constrained realization'' method that improves signal-to-noise by a factor of
  $\sim 2$ over a commonly-used aperture photometry technique. Measurements of the integrated tSZ
  $y$-parameter are used to break the kSZ velocity-optical depth degeneracy, and the effects of
  including CMB polarization and SZ profile uncertainties are also considered. A combination of
  future Stage 4 experiments should be able to measure the product of the growth and expansion
  rates, $\alpha\equiv f H$, to better than $1\%$ in bins of $\Delta z = 0.1$ out to
  $z \approx 1$ -- competitive with contemporary redshift-space distortion constraints from galaxy
  surveys.
\end{abstract}

  \date{\today}
  \maketitle

%% Section Introduction
\section{Introduction}\label{sec:intro}
  Galaxies and their big sisters, clusters, are test particles buffeted around by the
  cosmic gravitational field. If we could accurately measure their motions, as well as 
  their positions, it would be possible to learn much more about the origin and evolution
  of large scale structure, the fundamental properties of gravity, and the constituents
  of the Universe. Measurements of large scale flows, or peculiar velocities, are
  complementary to other approaches to mapping out the Universe that use, for example,
  the cosmic microwave background, the distribution of galaxies, and weak
  gravitational lensing.    
 
  For the past few decades, there have been numerous attempts to embark on this somewhat quixotic
  enterprise. There are now peculiar velocity catalogues with between $10^3 - 10^4$
  objects, some of which span the whole celestial sphere, others that are deeper and
  more targeted \cite{2001ApJ...546..681T,2003AJ....126.2268W,2007ApJS..172..599S,
  1999ApJ...510L..11D,2001MNRAS.321..277C,1999ApJ...522..647W,2014MNRAS.444.3926J}. It has
  been an arduous endeavour which, in some cases, has led to
  controversial results. Attempts at using direct distance indicators to galaxies or
  clusters (such as Tully-Fisher or fundamental plane relations) lead to shallow surveys
  with large uncertainties. Type Ia supernovae supply tighter constraints and allow
  for deeper surveys, but such surveys are, as yet, too sparse \cite{1995ApJ...445L..91R,
  2012MNRAS.420..447T};
  the same can be said of current kinetic Sunyaev-Zel'dovich measurements (the method that
  we will explore in this paper). On occasion, peculiar velocity surveys have
  led to results that are outliers within the standard cosmological canon: in the late
  $1980$s they were used to argue for an $\Omega\sim1$ universe \cite{1993ApJ...412....1D}, 
  while in the $1990$s and $2000$s they
  were used to claim evidence for excessive bulk motion on large scales 
  \cite{1994ApJ...425..418L,2009MNRAS.392..743W}. Given all this,
  and the rise of redshift space distortions (RSD) as a tool to learn about infall, direct
  measurements of peculiar velocities have become a neglected (and often maligned)
  area of research.
 
  This is about to change. We are embarking on a new era of cosmological surveys in which
  we will map out the Universe with unprecedented precision. In particular, by mapping
  the CMB over vast swathes of sky with fine resolution and high sensitivity,
  it should be possible to construct a completely new class of peculiar
  velocity catalogues that may revolutionize the field. By measuring the
  scattering of CMB photons off clusters, it is possible to pick up an effect -- the
  kinetic Sunyaev-Zel'dovich (kSZ) effect -- which is colour blind (i.e. follows the CMB
  blackbody spectrum), and proportional to the cluster peculiar velocity 
  \cite{1972CoASP...4..173S,1980ARA&A..18..537S}. This
  {\it direct} measurement of the peculiar velocity is independent of distance and
  redshift, and will allow us to construct deep surveys of the large scale flows of
  the Universe. 
 
  The kSZ effect has already been detected statistically, arguably using the WMAP data
  \cite{2008ApJ...686L..49K}, but most decisively with data from both the ACT
  \cite{Hand:2012ui}, Planck \cite{Ade:2015lza} and
  ACTPOL \cite{Schaan:2015uaa} experiments. Pointed (i.e. single-cluster) detections 
  also exist, e.g. \citep{2013ApJ...778...52S}. The
  significance of the detections is still poor and not good enough to be able
  to extract cosmological information, but the outlook is promising. A number of
  experiments have ramped up their sensitivity and scope, most notably
  Advanced ACT and SPT-3G \cite{2014SPIE.9153E..1PB}, and plans are under way to develop
  a consortium of telescopes, known as ``Stage 4" (S4), that will allow us to construct
  definitive catalogues of kSZ peculiar velocity constraints with $\mathcal{O}(10^4-10^5)$
  objects.
  
  There have been a number of attempts at forecasting what might be possible with future kSZ
  catalogues \cite{HernandezMonteagudo:2005ys,Bhattacharya:2006ke,Bhattacharya:2007sk,
  Mueller:2014nsa}. Indeed, using such catalogues to constrain the pairwise streaming
  velocity or the velocity correlation tensor seems promising, leading to improvements by
  factors of up to a few in the dark energy figure of merit. These statistics
  probe larger scales, less contaminated by non-linear growth and bias, and are complementary
  to more widely-used clustering statistics in redshift space.
 
  Even more promising is the idea of matching kSZ catalogues with density catalogues
  in such a way as to ``divide out'' the cosmic variance in the density/velocity field.
  It should be possible to reconstruct the most likely velocity field
  from a measurement of the 3D density field as traced by the number density of galaxies; 
  one can then compare the reconstructed velocity field with the kSZ measurements and
  find constraints on a combination of the growth rate of structure and the cluster
  optical depth/ionization fraction. Adding in other measurements, it may even be possible
  to disentangle the two -- making it possible to separately constrain cluster gas physics 
  and the linear growth rate. The purpose of this paper is to explore this
  approach, unpacking the different steps that go into such an estimation, and
  assessing the various alternatives at each step. Crucial to our analysis is a
  realistic assessment of the uncertainties that should be ascribed to this method.
 
  We structure the paper as follows. In Section \ref{sec:method} we describe the
  methods proposed to estimate the different ingredients of this procedure (the kSZ
  signal, the cluster optical depth, and the reconstructed velocities), as well as
  the forecasting formalism used. In Section \ref{sec:results} we compare
  three different kSZ measurement methods, and present the forecast constraints
  on the combination $\alpha\sim f H$ for each of them for several choices of current and
  next-generation CMB experiments and redshift surveys. Finally, in Section
  \ref{sec:discussion} we summarize the results and discuss the advantages and
  limitations of the proposed approach.

%-------------------------------------------------------------------------------  
\section{Growth reconstruction method}\label{sec:method}

  The idea behind the method explored here is to match a reconstructed velocity field with
  CMB measurements of the kSZ effect to obtain a per-source measurement of the growth rate of
  structure. The potential of combining kSZ measurements with galaxy surveys has been discussed
  before: forecasts for combinations of upcoming experiments were explored in
  \cite{Dore:2003ex,HernandezMonteagudo:2005ys,DeDeo:2005yr,Hill:2016dta}, and redshift surveys
  were essential in the first determination of the kSZ streaming velocity \cite{Hand:2012ui},
  as well as more recent attempts using the CMASS survey to pull out the kSZ signal at redshifts
  $z\sim0.4-0.7$ \cite{Schaan:2015uaa}. In this section we build on previous work and lay out,
  in detail, the observables that we need to work with, and the various steps involved in building
  up a reliable estimator for the growth rate.

  The fractional temperature fluctuations due to the thermal and kinetic Sunyaev-Zel'dovich effects
  are \citep{1980ARA&A..18..537S}
  \begin{align}\nonumber
    \left . \frac{\Delta {\rm T}}{{\rm T}} \right |_{\rm tSZ}(\nu,\nv)&=f_{\rm tSZ}(\nu)
      \frac{\sigma_T}{m_ec^2}\int P_e(l_z,\nv)\, dl_z\\\label{eq:tsz}
      &\equiv f_{\rm tSZ}(\nu) \, y(\nv) \\
    \left.\frac{\Delta {\rm T}}{{\rm T}} \right |_{\rm kSZ}(\nv)&=
      -\sigma_T \int ({{\bm \beta}\cdot\nv})\, n_e(l_z,\nv)\, dl_z \nonumber \\
      &\equiv -\beta_r \, \tau(\nv), \label{eq:ksz}
  \end{align}
  where $n_e$ and $P_e = k_{\rm B} n_e T_e$ are the electron number density and pressure,
  $\sigma_T$ is the Thomson scattering cross-section, and $\beta_r\equiv {\bf v}\cdot\nv/c$
  is the cluster's bulk velocity along the line of sight from the observer (parametrised by
  $l_z$). The spectral dependence of the thermal-Sunyaev-Zel'dovich (tSZ) effect is given by
  $f_{\rm tSZ}(\nu)$, and we
  have also defined the dimensionless Compton-$y$ parameter, $y(\theta)$, and optical depth,
  $\tau(\theta)$, profiles as a function of angle from the centre of the cluster (i.e.
  assuming sphericity). From Eq.~(\ref{eq:ksz}), it is clear that a detection of the kSZ
  effect corresponds to a measurement of the combination $\beta_r\times\tau$; if an external
  estimate of $\tau$ can be made, this determines the local velocity
  field.
  
  Let us now assume that we have a spectroscopic galaxy survey covering a volume that contains
  a number of tSZ-detected clusters, and that the redshifts of those clusters are known. As
  we will describe in Section \ref{ssec:recon}, the galaxy distribution can be used to
  reconstruct the velocity field at the cluster positions up to a factor
  \begin{equation}\label{eq:alpha1}
    \alpha(z)\equiv\frac{H(z)f(z)}{H_{\rm fid}(z)f_{\rm fid}(z)},
  \end{equation}
  where $H$ and $f$ are the expansion and growth rates, and the subscript ``fid'' labels
  quantities computed assuming the fiducial cosmology used to carry out the velocity
  reconstruction. For a given cosmology, the expected amplitude of the kSZ effect of a
  cluster $i$, as defined in Appendix~\ref{app:sz_models}, is $a^i_{\rm kSZ} =
  \beta_r^i \tau^i_{500}$. Assuming a value for $\tau_{500}$
  and an estimate of the cluster's radial velocity, $\hat{\beta}_r$, from the velocity field
  reconstruction, we can sum over all clusters in a redshift interval $[z, z +
  \delta z]$ to obtain a likelihood for $\alpha$,
  \begin{equation}
    -\log\mathcal{L}\equiv\chi^2(\alpha)=
    \sum_i\frac{\left(\alpha\hat{\beta}_r^i\tau_{500}^i-a^i_{\rm kSZ}\right)^2}
    {\mathcal{E}_i^2}.
  \end{equation}
  Here, $a_{\rm kSZ}$ is the measured value of the kSZ amplitude, and $\mathcal{E}$
  is the combined uncertainty in $\hat{\beta}_r$, $\tau_{500}$, and $a_{\rm kSZ}$ for each
  cluster. Assuming that the errors on these parameters are independent and
  Gaussian-distributed, the uncertainty on $\alpha$ is given by
  \begin{align}
   \sigma^{-2}_\alpha &=\sum_i \mathcal{E}_i^{-2} \label{eq:uncertainty_discrete}\\
   &\equiv \sum_i\left( \varepsilon_{a_{\rm kSZ},i}^2 
                      + \varepsilon_{\tau_{500},i}^2
                      + \varepsilon^2_{\beta_r,i}
                      + \varepsilon^2_{\tau_{500},i}\varepsilon^2_{\beta_r,i}\right)^{-1},
                      \nonumber
  \end{align}
  where $\varepsilon_x=\sigma_x/x$ are the relative uncertainties on the other three quantities.

  The final uncertainty on $\alpha$ for a given combination of CMB experiment and spectroscopic
  survey depends on the number of clusters for which this process can be carried out. 
  Both this, and the error on the measurement for each cluster, depend
  on the cluster halo mass, velocity, and redshift, and so we can rewrite
  Eq.~(\ref{eq:uncertainty_discrete}) as an integral over their expected distributions,
  \begin{align}\label{eq:errors_alpha}
      \sigma^{-2}_\alpha = &\,4\pi\,f_{\rm sky}\,\frac{r^2(z)\delta z}{H(z)}\\
      &\times \int_0^\infty dM\,
      \int_{-\infty}^\infty d\beta_r\frac{\tilde{\chi}(M,z)\,n(M,z)\,p(\beta_r|M,z)}
      {\mathcal{E}^2(M,\beta_r,z)}, \nonumber
  \end{align}  
  where $n(M,z)$ is the halo mass function (number density of dark matter haloes of mass
  $M\in[M,M+dM]$ in a given redshift interval), $p(\beta_r|M,z)$ is the distribution of
  halo radial velocities, and $\tilde{\chi}(M,z)$ is the detection efficiency for a cluster
  of a given mass and redshift for a given CMB experiment. The prefactor gives the volume of
  the redshift bin containing the clusters. For a given cosmology and set of survey
  specifications, we can therefore estimate the error on $\alpha$ by evaluating
  Eq.~(\ref{eq:errors_alpha}).
  
  In what follows, we model the various measurement uncertainty terms in
  $\mathcal{E}(M,\beta_r,z)$ (Sections \ref{ssec:method_ksz} and \ref{ssec:recon}) and the
  detection efficiency $\tilde{\chi}(M, z)$ for tSZ-selected clusters
  (Sect.~\ref{ssec:tsz_clusters}).

%-------------------------------------------------------------------------------
  \subsection{Cluster kSZ signal extraction}\label{ssec:method_ksz}
    Unlike the tSZ effect, which has a distinctive spectral dependence, the kSZ effect has
    the same flat spectrum as the primary CMB -- making the CMB anisotropies themselves an
    important source of contamination. Most kSZ detection methods therefore attempt to separate
    the two components by using differences in their angular distributions on the sky; while
    the angular extent of a typical galaxy cluster is of the order a few arcminutes
    (corresponding to $\ell \sim 3000$), the primary CMB anisotropies are strongly damped for
    $\ell \gtrsim 3000$, while dominating the power on much larger scales. An
    appropriately-designed angular high-pass filter can therefore be used to separate the two
    contributions.
    This is the idea behind most kSZ extraction methods (e.g. see 
    \citep{2004A&A...420...49F, 2014A&A...561A..97P} and references therein). We compare 
    three in this paper:
    \begin{itemize}
      \item The simplest is the {\it aperture photometry} (AP) filter, a blind method that
            uses a compensated circular filter with a radius similar to the cluster size to
            filter out the longer-wavelength CMB modes (Sect.~\ref{sssec:ap}).
      \item An enhanced semi-blind method, new to this work, that reconstructs and
            subtracts the CMB behind the aperture by using phase information from the
            surrounding area of sky, a technique known as {\it constrained realization} or
            in-painting (Sect.~\ref{sssec:semi_blind}).
      \item An optimal, minimum-variance {\it matched filter} estimator can be constructed
            by assuming a model for the spatial tSZ/kSZ profiles of the cluster. This entails
            making strong assumptions about the forms of $y(\theta)$ and $\tau(\theta)$,
            which leads to efficient filtering but potentially biased kSZ amplitude
            measurements (Sect.~\ref{sssec:matched}).
    \end{itemize}

%-------------------------------------------------------------------------------
    \subsubsection{Matched filtering}\label{sssec:matched}
      Matched filtering \citep[e.g.][]{2002MNRAS.336.1057H, 2006A&A...459..341M,
      2014MNRAS.443.2311L} entails specifying a model for the spatial and spectral variation
      of the tSZ and kSZ signals, and then convolving the resulting set of filters with the
      (foreground-cleaned) maps. A perfectly matched filter will recover an unbiased estimate
      of the SZ amplitudes by strongly suppresssing all other components with different
      spatial/spectral distributions. We model the data in each frequency band $\nu$ as
      \begin{equation}\label{eq:model1}
        m_\nu(\nv) = \sum_i {\rm U}^i_\nu(\nv)\cdot{\bf a}_i + {n}_\nu(\nv),
      \end{equation}
      where $m_\nu(\nv)\equiv \{ \Delta T(\nu,\nv) \}$ is the sky temperature measured in
      direction $\nv$, the noise term ${n}_\nu$ contains CMB, residual foreground (assumed
      zero here), and instrumental noise contributions, and the sum is over all clusters
      in the map. The matrix operator ${\rm U}_\nu(\nv) \equiv\left(u_{\rm tSZ}(\nu,\nv),
      u_{\rm kSZ}(\nu,\nv)\right)$ contains the tSZ and kSZ cluster spatial
      templates for each band, and ${\bf a} \equiv(a_{\rm tSZ}, a_{\rm kSZ})$ is a vector
      of amplitude parameters (see Appendix~\ref{app:sz_models} for definitions and
      parametric profile models).

      Assuming that the noise term is homogeneous, isotropic, and
      Gaussian-distributed\footnote{Note that these assumptions can easily be relaxed in a
      practical situation}, the log-likelihood for the amplitude parameters ${\bf a}$ is
      \begin{equation}
        \chi^2=\int d^2l \,\left[{\bf m}_{\bf l}-{\rm {\bf U}}_{\bf l}\cdot{\bf a}\right]^T
        \cdot {\rm {\bf C}}^{-1}_{\rm N}(l)\cdot
        \left[{\bf m}_{\bf l}-{\rm {\bf U}}_{\bf l}\cdot{\bf a}\right].
      \end{equation}
      where ${\bf l}$ labels the flat-sky Fourier modes of the $\nv$-dependent quantities in
      the previous equations, and the various bold quantities are appropriately-constructed
      block vectors and matrices containing the corresponding values for each Fourier
      mode/band/cluster. The total noise angular power spectrum is given by
      ${\rm {\bf C}}_{\rm N}(l)$.

      A minimum-variance estimate for ${\bf a}$ is then
      \begin{equation}
        \tilde{\bf a}\equiv {\rm Cov}(\tilde{\bf a}) \cdot
        \int d^2l\,{\rm {\bf U}}^T_{\bf l} {\rm {\bf C}}^{-1}_{\rm N}(l)\,
        {\bf m}_{\bf l},
      \end{equation}
      with covariance
      \begin{equation}\label{eq:matched_covariance}
        \left[{\rm Cov}(\tilde{\bf a})\right]^{-1} =
        \int d^2l\,{\rm {\bf U}}^T_{\bf l}{\rm {\bf C}}^{-1}_{\rm N}(l){\rm {\bf U}}_{\bf l}.
      \end{equation}
      As stated previously, we assume that the only relevant noise components are the primary
      CMB anisotropies and instrumental noise. The noise power spectrum is then
      \begin{equation}
        [{\rm C}_{\rm N}(l)]_{\nu\nu^\prime} = {\rm C}^{\rm CMB}_l +
        \frac{N^\nu_l}{(B_l^\nu)^2}\delta_{\nu \nu^\prime},
      \end{equation}
      where $N^\nu_l$ and $B_l^\nu$ are the noise power spectrum and harmonic coefficients
      of the instrumental beam profile in frequency channel $\nu$. In our fiducial analysis
      we will assume uncorrelated noise, so that $N^\nu_l = \sigma_{{\rm N}, \nu}^2$, where
      $\sigma_{{\rm N}, \nu}$ is the rms noise per steradian in each channel. Note that
      correlated instrumental noise (e.g. due to coherent atmospheric fluctuations) is
      expected to be non-negligible for actual ground-based experiments.

      While matched filtering yields a minimum variance estimate of the kSZ amplitude, its
      effectiveness depends upon selecting the correct SZ profiles; otherwise, the estimates
      will be biased. Clusters are far from simple, ideal objects, however -- profiles vary
      significantly between clusters, and the parametric profiles that are typically used
      tend to give only approximate fits to any given object. One could marginalize over
      the profile parameters, imposing a prior on them based on hydrodynamic simulations,
      for example, but even state of the art simulations fail to fit some features of real
      cluster samples. Matched filtering therefore necessitates a strong (and potentially
      unrealistic) prior to be placed on cluster physics, so substantial care must be
      exercised in the use of this technique.

%-------------------------------------------------------------------------------
    \subsubsection{Constrained realizations}
    \label{sssec:semi_blind}

      While the exact shape of the mean kSZ cluster profile is currently very uncertain, we
      have precise information about the statistics of the primary CMB -- its temperature power
      spectrum is modelled, and well-measured, out to high $\ell$. This information can be used
      to construct and subtract a maximum-likelihood (ML) estimate of the CMB {\it behind} the
      cluster, without needing to assume a specific cluster model. The method for doing this,
      called constrained realization (CR) or `in-painting' of the CMB, has been used previously
      to fill-in masked regions of CMB maps, for example
      \citep[e.g.][]{2008StMet...5..289A, 2008PhRvD..77l3539I, 2012MNRAS.424.1694B,
      2012ApJ...750L...9K, 2013A&A...549A.111E, 2014ApJS..210...24S}.

      Begin by assuming that a cluster catalogue has been obtained, and a tSZ- and foreground-free
      map has been produced using a frequency-dependent filtering scheme. Our CR method then
      proceeds as follows:
      \begin{enumerate}
        \item Define a disc ${\cal D}$ of radius $\theta_R$ around the centre of each 
              cluster that is large enough to encompass the bulk of the cluster's kSZ 
              emission.
        \item Use the measured CMB fluctuations outside the disc to infer the
              ML value inside the disc (as described below).
        \item Subtract the maximum-likelihood estimate from inside the disc, and 
              integrate the residual in the disc area to estimate the total kSZ flux.
      \end{enumerate}
      The ML CMB temperature field, $\bar{T}_{\textrm{CMB}}$, can be obtained by Wiener-filtering
      the (cleaned) map with the disc region ${\cal D}$ masked out \footnote{The Wiener-filter
      solution covers the full map, including the masked region.}. The covariance of the residual
      CMB field, $T^{\rm (true)}_{\textrm{CMB}}-\bar{T}_{\textrm{CMB}}$, is given by
      $\left(\bm{C}^{-1}+\bm{N}^{-1} \right)^{-1}$ \cite{2015ApJS..219...10B}, where $\bm{C}$ is
      the CMB covariance matrix (fixed to a best-fit power spectrum model), and $\bm{N}$ is the
      noise covariance matrix assuming infinite noise inside the disc,
      \begin{equation}
        \bm{N}^{-1}_{ij}=\left\{
        \begin{array}{ll}
          \sigma_{\rm pix}^{-2}\delta_{ij} &  \\
          0 &  \mbox{if $i,j \in {\cal D}$ }
        \end{array}
        \right.,
      \end{equation}
      where $\sigma_{\rm pix}^2$ is the per-pixel noise variance (assumed homogeneous) of the
      tSZ-cleaned map outside the masked disc region. Our estimator for the kSZ flux is then
      \begin{equation}
        \hat{a}^{\rm CR}_{\rm kSZ}=
        \sum_{i \in {\cal D}} (m_{i} - \bar{T}_{i,\textrm{CMB}}) \,\Omega_{\rm pix}
      \end{equation} 
      where $m_i$ is the value of the tSZ-filtered map in pixel $i$ (with pixel area
      $\Omega_{\rm pix}$), and the sum is over all pixels inside $\mathcal{D}$.
      The variance of $\hat{a}^{\rm CR}_{\rm kSZ}$ is then given by
      \begin{equation}\label{eq:ci_variance}
        {\rm Var} (a^{\rm CR}_{\rm kSZ}) =  \sum_{i,j \in {\cal D}}
        \left[ \left(\bm{C}^{-1}+\bm{N}^{-1} \right)_{ij}^{-1} + \sigma^{2}_{\rm pix}
        \delta_{ij} \right] \Omega_{\rm pix}^2.
      \end{equation}
      The first term in square brackets is the variance for the reconstructed CMB from above,
      and the second is the instrumental noise variance (which would affect the kSZ term even
      in the absence of the CMB). We have assumed that the effect of the tSZ and foreground
      components is fully encapsulated in the enlarged noise variance of the tSZ-cleaned map.

      The first term in Eq. \ref{eq:ci_variance} is difficult to evaluate, so we compute it
      as
      \begin{equation}
        \sum_{i,j\in{\cal D}}\left(\bm{C}^{-1}+\bm{N}^{-1} \right)_{ij}^{-1}=
        {\bf u}^T\cdot\left(\bm{C}^{-1}+\bm{N}^{-1} \right)^{-1}\cdot{\bf u},
      \end{equation}
      where ${\bf u}$ is a vector containing $1$ in all pixels inside the disc, and $0$
      otherwise. The matrix inversion $(\bm{C}^{-1} + \bm{N}^{-1})^{-1}\cdot{\bf u}$ is
      carried out using a preconditioned conjugate gradient solver, with
      $\bm{C}^{-1}$ and $\bm{N}^{-1}$ applied in Fourier and real space respectively (where each
      matrix is sparsest) \footnote{Note that several other fast methods for solving the Wiener
      filtering equation have been developed in the literature, e.g.
      \citep{2004ApJS..155..227E, 2012MNRAS.424.1694B, 2013A&A...549A.111E,
      2014ApJS..210...24S}.}. We checked that this estimate of the uncertainty was independent
      of pixel size.

      This method has one free parameter: the choice of disc radius,
      $\theta_R$. For this work, we chose $\theta_R$ to be such that $80\%$ of the
      (expected) beam-convolved kSZ signal was enclosed in the disc. (The same criterion was
      used for the aperture photometry filter described in the next section.) The resulting flux
      estimate will therefore be biased, as some fraction of the signal will fall outside the
      disc; this bias must be corrected for analytically, or using simulations. A real analysis
      would presumably also compare several choices of $\theta_R$ to ensure stability of the
      results \citep[c.f.][]{Schaan:2015uaa}.

      Because it does not use information about the shape of the SZ profiles, the performance
      of this estimator is dictated by the noise level and the size of the disc that we
      consider. The CMB correlation function drops rapidly with separation angle, making
      the uncertainty on the residual $T_{\textrm{CMB}}-\bar{T}_{\textrm{CMB}}$ a steep
      function of the disc radius, $\theta_R$. The uncertainty for clusters that subtend
      larger angles is therefore dominated by the CMB fluctuations, while
      the instrumental noise becomes more relevant for smaller discs. 

%-------------------------------------------------------------------------------
    \subsubsection{Aperture photometry filter}\label{sssec:ap}
      Aperture photometry (AP) attempts to avoid specific assumptions about both the CMB
      statistics and cluster properties. It is conceptually similar to the constrained realization
      method from above, in that it tries to estimate and subtract the CMB fluctuations inside 
      a disc centred around each cluster. Its `modelling' of the CMB is much simpler, however.

      The method defines two concentric circles around each cluster, with radii $\theta_R$ and
      $\sqrt{2}\theta_R$ respectively, such that the areas of the inner and outer regions are the
      same. If CMB fluctuations have a typical angular size much larger than $\theta_R$, they
      will be almost constant over the aperture. Subtracting the flux integrated over the outer
      region from the inner region will therefore result in zero mean CMB signal. Assuming that
      $\theta_R$ has been chosen such that most of the kSZ flux is inside the inner region, the
      integral of the residual there will be a good estimate of the total kSZ flux.

      The simplicity of the AP method makes it possible to evaluate its performance analytically.
      The estimated kSZ contribution inside the inner region is
      \begin{equation}\nonumber
        \tilde{\Delta}_{\rm kSZ}(\theta,\phi)=m(\theta,\phi)-\frac{1}{\pi\theta_R^2}\int_0^{2\pi}
        d\phi'\int_{\theta_R}^{\sqrt{2}\theta_R}d\theta'\,\theta'\,m(\theta',\phi'),
      \end{equation}
      where $m(\nv)$ is the tSZ-cleaned map, and $(\theta,\phi)$ are cylindrical coordinates
      defined with respect to the centre of the aperture. The total kSZ flux in the inner region
      is
      \begin{align}\nonumber
        a^{\rm AP}_{\rm kSZ}&=\int_0^{2\pi}d\phi\int_0^{\theta_R}d\theta\,\theta\,
              \tilde{\Delta}_{\rm kSZ}(\theta,\phi)\\
            &=\int_0^{2\pi}d\phi\int_0^\infty d\theta\,\theta\,
              W_{\rm AP}(\theta|\theta_R)\,m(\theta,\phi),
      \end{align}
      where the AP window function $W_{\rm AP}(\theta|\theta_R)$ is $1$ for
      $0<\theta<\theta_R$, $-1$ for $\theta_R<\theta<\sqrt{2}\theta_R$, and $0$ otherwise. For
      homogeneous and isotropic noise, the variance of $a^{\rm AP}_{\rm kSZ}$ is
      \begin{equation}
        {\rm Var}(a^{\rm AP}_{\rm kSZ})=2\pi\,\theta_R^4\int_0^\infty dl\,l\,C_N(l)\,
        \left|\tilde{W}_{\rm AP}(l|\theta_R)\right|^2,
      \end{equation}
      where $C_N(l)$ is the noise power spectrum (including CMB and instrumental noise), and
      $\tilde{W}_{\rm AP}$ is the Fourier transform of the AP filter, given by
      \begin{equation}
        \tilde{W}_{\rm AP}(l|\theta_R)=\frac{2J_1(l\theta_R)-
        \sqrt{2}J_1(\sqrt{2}l\theta_R)}{l\theta_R}.
      \end{equation}
      We also validated this calculation numerically, using Gaussian realizations of the CMB.

      While this method is in some sense model-independent, it is also biased (like the CR
      method, above), and has higher variance. The latter is a consequence
      of the non-vanishing primary CMB power on scales of order the aperture size; while
      suppressed due to Silk damping, CMB anisotropies still dominate the kSZ signal on the
      typical (arcminute) angular scales of clusters. These contributions are not filtered by
      the AP method, and contribute significantly to the variance.

%-------------------------------------------------------------------------------
    \subsection{Cluster optical depth}\label{ssec:tau}
      As discussed above (see Eq.~\ref{eq:ksz}), the kSZ flux measures a degenerate combination of
      optical depth and velocity, and so additional information is needed to recover the
      velocities by themselves. This can be achieved through a number of different methods -- for
      example, the mean optical depth as a function of cluster mass and redshift can be calibrated
      using simulations \cite{2015ApJ...806...43B}, or from CMB polarization data
      \cite{1999MNRAS.310..765S}. One can also independently
      estimate $\tau$ by self-consistently modelling the ionized gas profile, or combining X-ray
      and tSZ information \cite{2005ApJ...635...22S}.

      Following the results of \cite{battaglia_inprep} using hydrodynamical
      simulations, we have assumed a log-normal scaling relation between the mean optical depth
      and the integrated Compton-$y$ parameter:
      \begin{equation}
        \log_{10}\bar{\tau}_{500}=A+B\log_{10}\bar{Y}_{500},
      \end{equation}
      and that the value of $\tau_{500}$ for individual clusters will be scattered around this
      relation with a dispersion $\Delta\tau/\tau$. The relative uncertainty on $\tau_{500}$ is then 
      \begin{equation}\label{eq:error_tau}
        \varepsilon_\tau=\sqrt{B^2\left(\frac{\sigma_Y}{Y_{500}}\right)^2+
        \left(\frac{\Delta\tau}{\tau}\right)^2},
      \end{equation}
      where $\sigma_Y$ is the statistical uncertainty in the measurement of $Y_{500}$, given in
      Eq. \ref{eq:matched_covariance}, and we have assumed a scatter $\Delta\tau/\tau=0.15$, in
      agreement with simulations \cite{2015ApJ...806...43B}. Note that this value corresponds to
      the scatter in $\tau$ for a given mass range, and not the scatter around the
      $Y_{500}-\tau_{500}$ relation. In that sense Eq. \ref{eq:error_tau} would conservatively
      overestimate the total uncertainty on $\tau_{500}$.

      Regarding the scaling parameter, $B$, here we have used the scaling of $Y_{500}$
      and $\tau_{500}$ with halo mass, $M_{500}$, according to the cluster models described in
      Appendix~\ref{app:sz_models}, to obtain
      \begin{equation}
        B=\frac{\alpha-4/3}{\alpha-2/3} \approx 0.41,
      \end{equation}
      where $\alpha\simeq1.79$ is the scaling exponent of the $Y_{500}-M_{500}$ relation
      (Eq. \ref{eq:yscaling}).

%-------------------------------------------------------------------------------
  \subsection{Detection efficiency of SZ-selected clusters}\label{ssec:tsz_clusters}
    For a tSZ-selected cluster survey, the detection efficiency can be written as
    \begin{align}
      \tilde{\chi}(M_{500},z)=&\int d(\ln Y^{\rm true}_{500}) \int_{q\sigma_N}^\infty
      dY_{500}^{\rm obs} \label{eq:chi_sz1}\\
      &~~~~~ P_{\rm SZ}(\ln Y^{\rm true}_{500}|M_{500},z)\,
      P_{\rm det}(Y_{500}^{\rm obs}|Y_{500}^{\rm true}), \nonumber
    \end{align}
    where $P_{\rm det}$ is the probability of obtaining a measurement $Y_{500}^{\rm obs}$
    for a true integrated tSZ flux $Y_{500}^{\rm true}$ (see Appendix \ref{app:sz_models}), and
    $P_{\rm SZ}$ is the distribution of integrated tSZ fluxes for clusters of mass $M_{500}$ at
    redshift $z$, which accounts for the intrinsic scatter in the $Y-M$ relation. We have
    assumed a detection threshold of $q \sigma_N$, where $\sigma_N$ is the noise on the
    measurement of $Y_{500}$ (given by Eq.~\ref{eq:matched_covariance} for matched-filter
    detections), and $q$ is the detection level above which clusters are accepted (e.g. $q=5$ 
    denotes a $5\sigma$ detection threshold).

    Assuming Gaussian errors on the tSZ flux, the inner integral in Eq.~(\ref{eq:chi_sz1}) is
    \begin{equation}\nonumber
      \int_{q\sigma_N}^\infty dY_{500}^{\rm obs}\,
      P_{\rm det}(Y_{500}^{\rm obs}| Y_{500}^{\rm true})
      =\frac{1}{2}\left[1+{\rm erf}\left[\frac{Y_{500}^{\rm true}-q\sigma_N}
      {\sqrt{2}\sigma_N}\right]\right].
    \end{equation}
    The distribution of true tSZ fluxes is usually assumed to take a log-normal form,
    \begin{equation}
      P_{\rm SZ}(\ln Y_{500}|M_{500},z)=\frac{1}{\sqrt{2\pi}\sigma_{\ln Y_{500}}}
      \exp\left[-\frac{\ln^2(Y_{500}/\bar{Y}_{500})}{2\,\sigma^2_{\ln Y_{500}}}\right],
    \end{equation}
    where $\bar{Y}_{500}(M_{500},z)$ and $\sigma_{\ln Y_{500}}$ are the mean and
    intrinsic scatter in the $Y-M$ relation. We adopt the empirical fitting function
    from \cite{Ade:2015fva}, given by
    \begin{equation}\label{eq:yscaling}
      \bar{Y}_{500}=Y_*
                \left[\frac{d_A(z)}{100\,{\rm Mpc}/h}\right]^{-2}
                \left[\frac{(1-b)M_{500}}{10^{14}\,M_\odot/h}\right]^\alpha
                E^\beta(z),
    \end{equation}
    where $d_A$ is the angular diameter distance, $1-b=0.8$,
    $Y_*=2.42\times10^{-10} {\rm sr}^2$, $\alpha=1.79\pm0.08$, $\beta=0.66\pm0.5$,
    $E(z)\equiv H(z)/H_0$, and $\sigma_{\ln Y_{500}}=0.127\pm0.023$.

%-------------------------------------------------------------------------------
  \subsection{Galaxy survey velocity reconstruction}\label{ssec:recon}
    In Newtonian theory, the relationship between the velocity and density fields is fully
    described by three non-linear equations: the continuity, Euler, and Poisson equations
    \cite{2008gady.book.....B}. The continuity equation reads
    \begin{equation}
      \dot{\delta}+\frac{1}{a}\nabla\cdot\left((1+\delta)\,{\bf v}\right)=0.
    \end{equation}
    While evaluating the time derivative $\dot{\delta}$ in general requires solving the
    non-linear system of equations in full, the density field grows self-similarly
    ($\delta(t,{\bf x})=D(t)\,\delta(t_0,{\bf x})$) in linear theory. After linearization,
    this allows us to rewrite the (Fourier space) continuity equation as
    \begin{equation}\label{eq:v_recon}
      {\bf v}(t,{\bf k})=\frac{H\,f}{a}\,\frac{i{\bf k}}{k^2}\delta(t,{\bf k}),
    \end{equation}
    where $a$ is the scale factor, and $H\equiv\dot{a}/a$ and $f\equiv\dot{D}/D$ are
    the expansion and growth rates. A measurement of the three-dimensional density field
    can therefore be used to infer the velocity field on linear scales. In practise,
    this can be achieved by using the
    number counts from a spectroscopic galaxy survey as a (biased) proxy for the true
    density. Several sources of systematic uncertainties must be addressed, however:

    \paragraph{Non-linearities:} Eq.~(\ref{eq:v_recon}) is only valid in the linear regime;
    non-linearities may introduce a bias in the recovered velocities. The impact of this
    effect can be mitigated by filtering out the smallest non-linear scales, at the cost
    of introducing extra variance in the reconstructed velocity field. We provide a more
    quantitative description of these effects below.

    \paragraph{Galaxy bias:} The relation between the observed galaxy number density and
    the true matter density field must be correctly modelled in order to avoid a biased
    reconstructed velocity field. While the connection between both fields has been shown
    to be well described by a linear, deterministic, and scale-independent bias factor,
    $\delta_{\rm gal}=b_g\,\delta$, on large scales, possible deviations from this model
    on small scales are a potentially dangerous systematic uncertainty.

    \paragraph{Shot noise:} Noise due to a low number density of detected galaxies can
    significantly increase the variance of the reconstructed velocity field. A 
    Wiener-filtering approach can be used to down-weight shot noise-dominated scales
    \cite{2016MNRAS.457.2068C}.

    \paragraph{Redshift-space distortions:} The non-zero radial peculiar velocities of
    galaxies modify their apparent redshift, and hence distort the recovered density
    field in an anisotropic manner. Redshift-space distortions are, however, well
    understood in linear theory, and can be fully incorporated into Eq. \ref{eq:v_recon}.
    An incorrect modelling of non-linear RSDs could introduce important systematics in
    the reconstructed velocity field, however.

    We obtained a best-case estimate of how accurately the true cluster velocities can
    realistically be recovered from the reconstructed velocity field by running a simple
    reconstruction algorithm on a suite of N-body simulations. The simulations were carried out
    using Gadget-2 \cite{Springel:2005mi}, a tree-PM gravitational solver, which was run
    on initial conditions generated using second-order Lagrangian perturbation theory
    \cite{2006MNRAS.373..369C} at $z=49$ \footnote{Code available at
    \href{https://github.com/damonge/IC_DAM}{https://github.com/damonge/IC\_DAM}.}.
    Each simulation contained $512^3$ dark matter particles in a box of size
    $L_{\rm box}=1400\,h^{-1}{\rm Mpc}$. A $\Lambda$CDM cosmological model was used, with
    parameters $(\Omega_M,\Omega_\Lambda,\Omega_b,h,\sigma_8,n_s)=
    (0.315,0.685,0.049,0.67,0.84,0.96)$, compatible with the latest constraints from Planck
    \cite{Ade:2015xua}. Snapshots were output at redshifts $z=0, 0.05, 0.15, 0.3$ and $1$,
    and dark matter halos found in each of them using a friends-of-friends
    algorithm \footnote{Code available at
    \href{https://github.com/damonge/MatchMaker}{https://github.com/damonge/MatchMaker}.} with
    linking length $b_{\rm FOF}=0.2$.

    For each snapshot, we estimated the reconstructed velocity for each halo as follows:
    \begin{enumerate}
      \item The density field is estimated on a Cartesian grid of size $N_{\rm grid}=512$
            using a Cloud-In-Cell algorithm.
      \item The density field is then smoothed using a Gaussian filter with standard
            deviation $R_G$ as the characteristic scale. We studied the dependence of
            the reconstructed velocity field on the choice of smoothing scale by repeating
            this step for $R_G=\{0,\,0.5,\,1,\,2,\,4,8\}\,h^{-1}{\rm Mpc}$.
      \item The velocity field is then estimated from the smoothed density field by solving
            the linearized continuity equation in Fourier space (Eq. \ref{eq:v_recon}).
      \item A reconstructed velocity is assigned to each halo by interpolating the
            velocity field to the halo position, using a trilinear interpolation scheme.
    \end{enumerate}
    We then compute the relative error between the reconstructed and true halo velocities
    for each halo, and study its statistics as a function of redshift and smoothing scale
    in different mass bins.

\begin{figure}[t]
      \centering
      \includegraphics[width=0.48\textwidth]{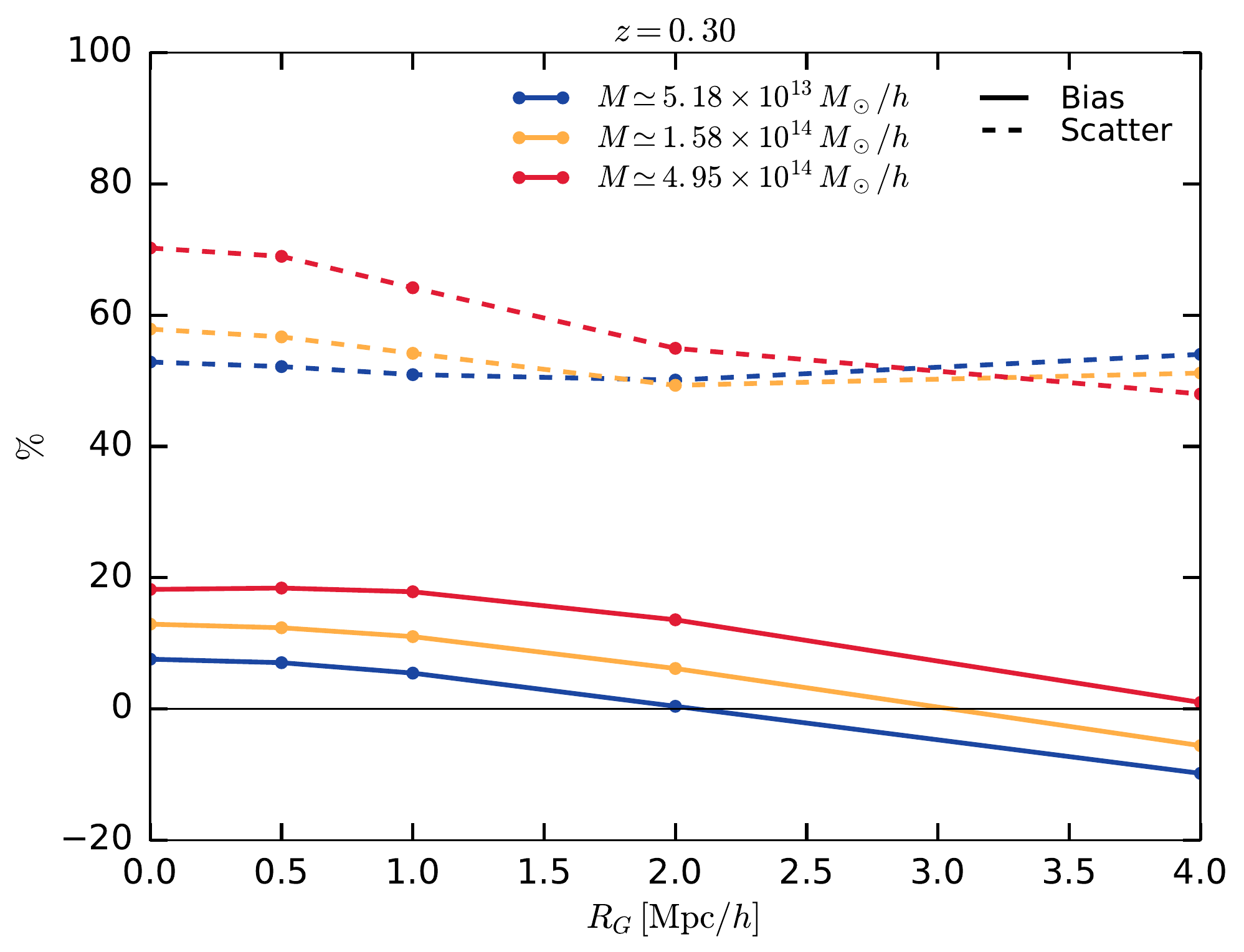}
      \caption{Relative bias (solid lines) and standard deviation (dashed lines) 
               of the reconstructed halo velocities for different Gaussian smoothing 
               scales, in three mass bins at $z=0.3$.}
      \label{fig:velrec}
      \vspace{-1em}
    \end{figure}

    For halo masses in the range of interest, we find that it is always possible to find
    a smoothing scale that yields an unbiased estimate of the halo velocity, as well as
    roughly attaining minimum variance. Fig.~\ref{fig:velrec} shows this explicitly for
    the $z=0.3$ snapshot. In all cases, we found the optimal smoothing scale to be in the
    range $R_G\in (2,6)\, h^{-1}{\rm Mpc}$. The relative error in the reconstructed radial
    velocities is $\sim 50\%$, and is well fit by
    \begin{equation}\label{eq:error_reconst}
      \varepsilon_{\beta_r}=\varepsilon_0\,(1+z)^{\alpha_0}+
      \varepsilon_1\,(1+z)^{\alpha_1}\,\log_{10}
      \left(\frac{M}{10^{14}\,h^{-1}\,M_\odot}\right),
    \end{equation}
    with $\varepsilon_0=0.50$, $\alpha_0=-0.01$, $\varepsilon_1=0.02$ and $\alpha_1=-1.9$.

    This estimate of the relative error due to the velocity reconstruction includes the
    contribution from non-linear scales, but none of the other three effects listed above
    (galaxy bias, shot noise, and RSDs). A thorough evaluation of these lies beyond the
    scope of this paper. In any case, as evidenced by the results shown in Section
    \ref{ssec:ksz_methods}, the uncertainty in the measured kSZ amplitude for each cluster
    should dominate the combined total uncertainty of the method
    (Eq.~\ref{eq:uncertainty_discrete}), so we do not expect these caveats to significantly
    affect our results.

    We also used the halo catalogues from these simulations to estimate the distribution of
    radial velocities $p(\beta_r|M,z)$ that enters Eq.~\ref{eq:errors_alpha}. We find that
    $v_r\equiv c\,\beta_r$ is approximately Gaussian-distributed, with zero mean and a
    standard deviation given by
    \begin{equation}
      \sigma_v(M,z)\simeq\sigma_0(1+z)^{\gamma_0}-\sigma_1(1+z)^{\gamma_1}\,
      \log_{10}\left(\frac{M}{10^{14}\,h^{-1}\,M_{\odot}}\right),
    \end{equation}
    with $(\sigma_0, \sigma_1)=(312, 22)\,{\rm km/s}$ and
    $(\gamma_0, \gamma_1)=(0.87, 1.05)$.

    Finally, note that even though we have so far claimed that this method is able
    to yield a measurement of the quantity $\alpha$ defined in Eq.~\ref{eq:alpha1}, since
    $f\,H$ is the combination entering Eq. \ref{eq:v_recon}, the reconstructed velocity field
    is also sensitive to the normalization of the matter density field $\delta$ estimated
    from the galaxy overdensity. This relation is, on linear scales, determined by the
    galaxy bias $b_g$ as well as the overall normalization of the density power spectrum
    which can be encoded in the parameter $\sigma_8$. Thus, in reality, this method
    measures the combination
    \begin{equation}
      \alpha\equiv\frac{f(z)\,H(z)\,b_g\,\sigma_8}
      {f_{\rm fid}\,H_{\rm fid}\,b_{g,{\rm fid}}\,\sigma_{8,{\rm fid}}}.
    \end{equation}
    It should be possible to obtain tight priors on $b_g$ and $\sigma_8$ from
    measurements of galaxy clustering and CMB power spectra, so we will
    regard $\alpha$ as mainly measuring the product $f\,H$ in what follows. The existing
    uncertainties on these quantities should, however, be borne in mind.

%-------------------------------------------------------------------------------
\section{Results}\label{sec:results}
  In this section, we forecast how well each of the three kSZ extraction methods will be
  able to measure the expansion and growth rates using forthcoming Stage 3 and 4 ground-based
  CMB experiments.
  
\begin{figure*}[t]
      \centering
      \includegraphics[width=0.49\textwidth]{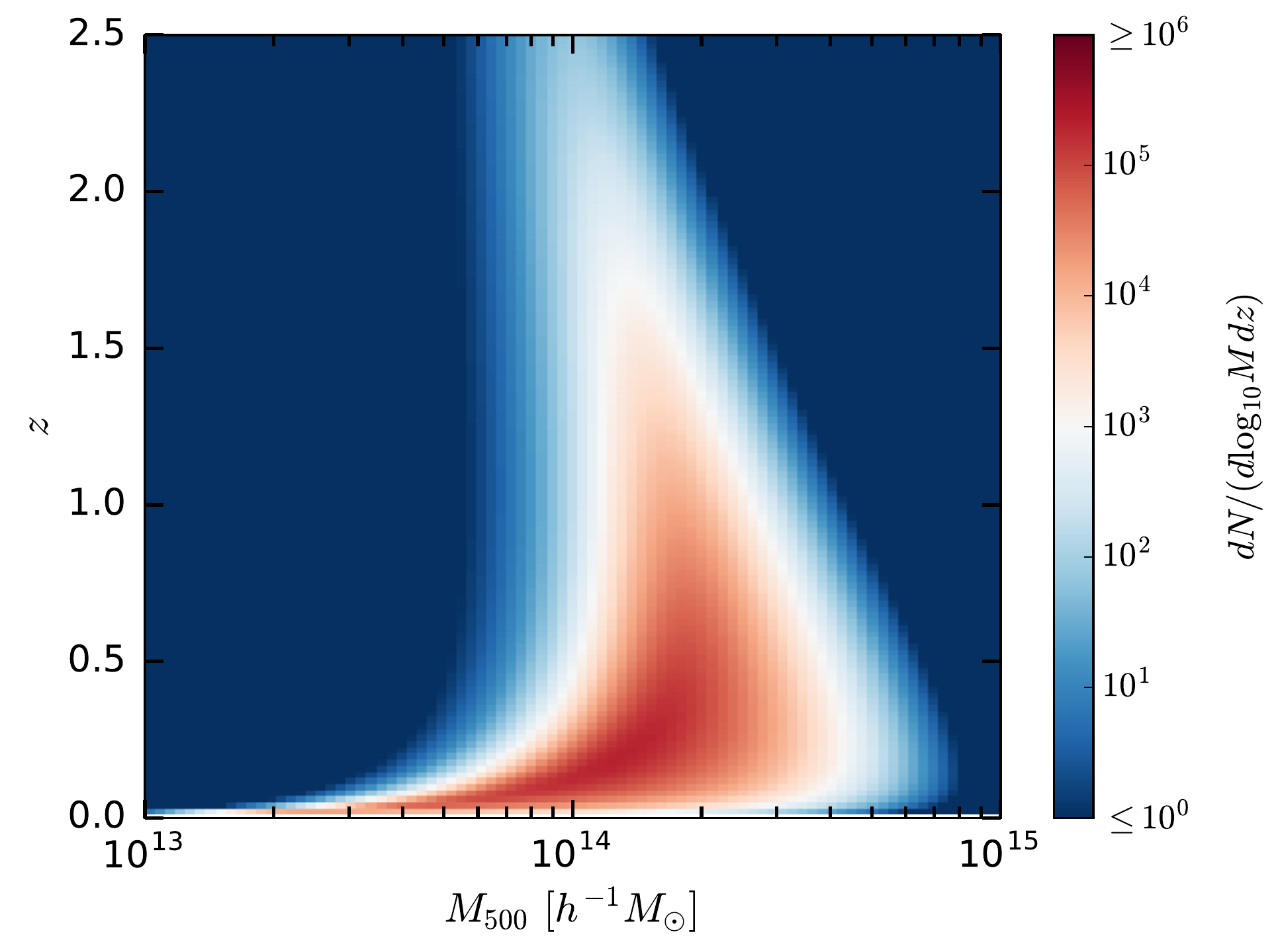}
      \includegraphics[width=0.49\textwidth]{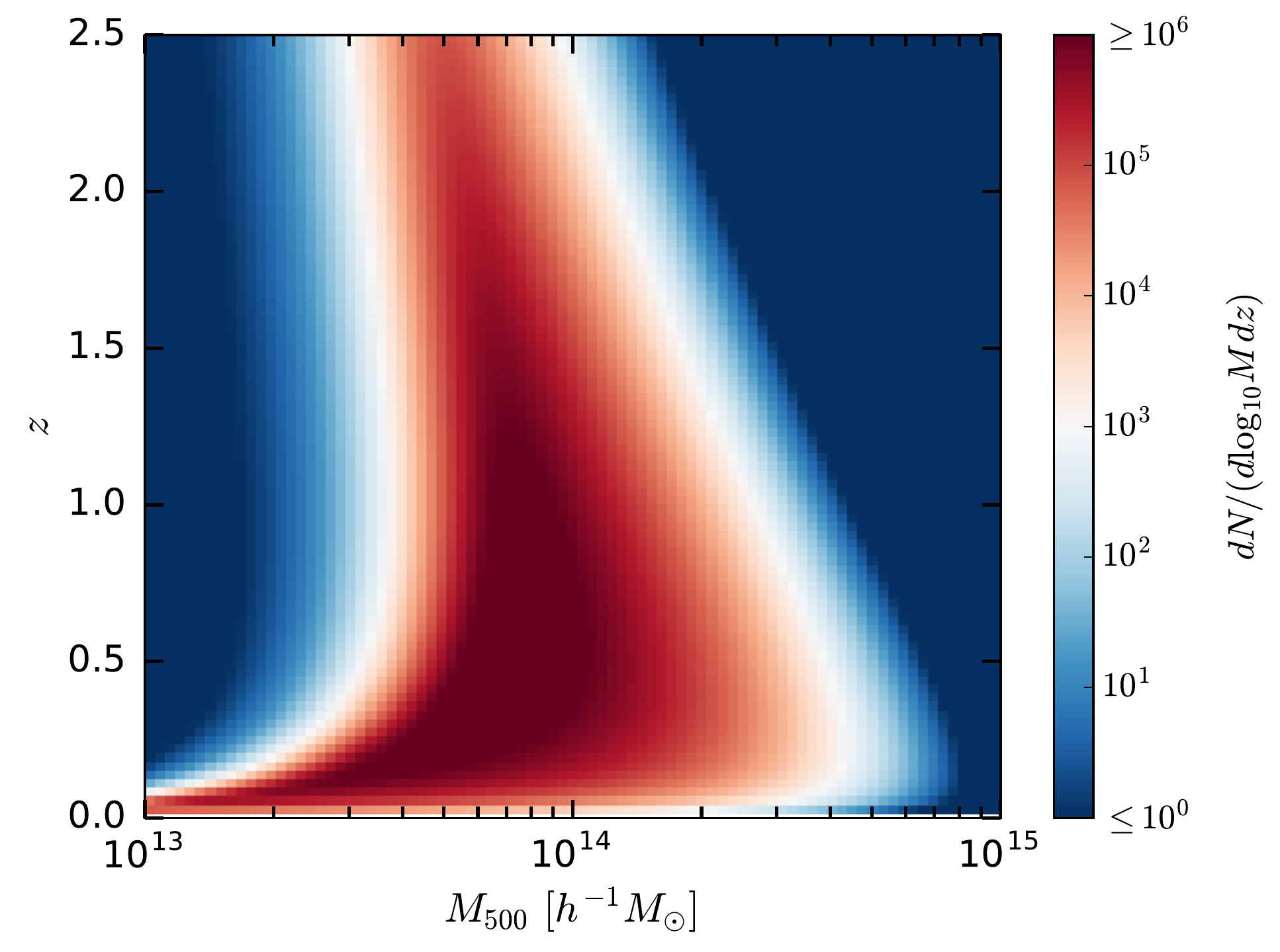}
      \caption{Expected mass and redshift distributions for tSZ-selected clusters detected
               with the S3 (left) and S4 (right) experiments.}
      \label{fig:cluster_distributions}
    \end{figure*}

%-------------------------------------------------------------------------------
  \subsection{Experimental setup}
    \begin{table}[b]
    \centering{
      \renewcommand*{\arraystretch}{1.2}
      \begin{tabular}{|c|C{1cm}C{1cm}|C{1cm}C{1cm}|}
        \hline
    Frequency & \multicolumn{2}{c|}{Noise RMS} & \multicolumn{2}{c|}{Beam FWHM} \\
    (GHz) & \multicolumn{2}{c|}{($\uKam$)} & \multicolumn{2}{c|}{(arcmin)} \\
    \hline
    & S3 & S4 & S3 & S4 \\
    \hline
    ~28 & 78.0  & 9.8 & 7.1 & 14.0  \\
    ~41 & 71.0  & 8.9 & 4.8 & 10.0  \\
    ~90 & ~7.8   & 1.0 & 2.2 & ~5.0   \\
    150 & ~6.9   & 0.9 & 1.3 & ~2.8 \\
    230 & 25.0  & 3.1 & 0.9 & ~2.0 \\
    \hline
    \end{tabular}}
    \caption{Specifications for representative CMB experiments.}
    \label{tab:cmbexp}
    \end{table}

    The current state of the art in CMB observation combines datasets from full-sky,
    space-based experiments (WMAP \cite{2013ApJS..208...20B} and Planck
    \cite{2014A&A...571A...1P}) with ``Stage 2'' ground-based experiments that focus on
    mapping the small-scale CMB anisotropies (e.g. ACTPol \cite{2014JCAP...10..007N}, SPT-Pol
    \cite{2015ApJ...807..151K}, and POLARBEAR \cite{2014ApJ...794..171T}). Over the next few
    years, enhanced Stage 3 (S3) ground-based experiments (e.g. AdvACT \cite{2015arXiv151002809H}
    and SPT-3G \cite{2014SPIE.9153E..1PB})
    will be rolled out, with larger numbers of detectors, multiple frequency channels, and the
    ability to survey a larger fraction of the sky. The high angular resolution and low noise
    levels of these experiments will make them ideal for cluster science, producing SZ catalogues
    that contain $\mathcal{O}(10^4)$ sources over a wide range of masses and redshifts.

    S3 experiments will eventually be superseded by a Stage 4 (S4) experiment, possibly
    composed of a set of ground-based facilities. Such an experiment would cover
    $\sim\!20,000\,{\rm deg}^2$ on the sky, with noise levels of around $1\uKam$. Such high
    sensitivity and large sky coverage is expected to increase the size of the corresponding
    cluster catalogue by at least an order of magnitude, making S4 an ideal experiment for the
    application of the method described here.

    We consider a representative experimental specification for each Stage. For S3, we assume
    a wide ($f_{\rm sky}=0.4$) survey with characteristics similar to those of AdvACT. The
    likely design of S4 is much less certain, so we consider an enhanced version of the S3
    setup, with twice the beam width, eight times the sensitivity, and the same sky fraction.
    We assume Gaussian beams in every band for S3 and S4. The specifications for both
    experiments are detailed in Table \ref{tab:cmbexp}.

\begin{figure}[t]
      \centering
      \includegraphics[width=0.45\textwidth]{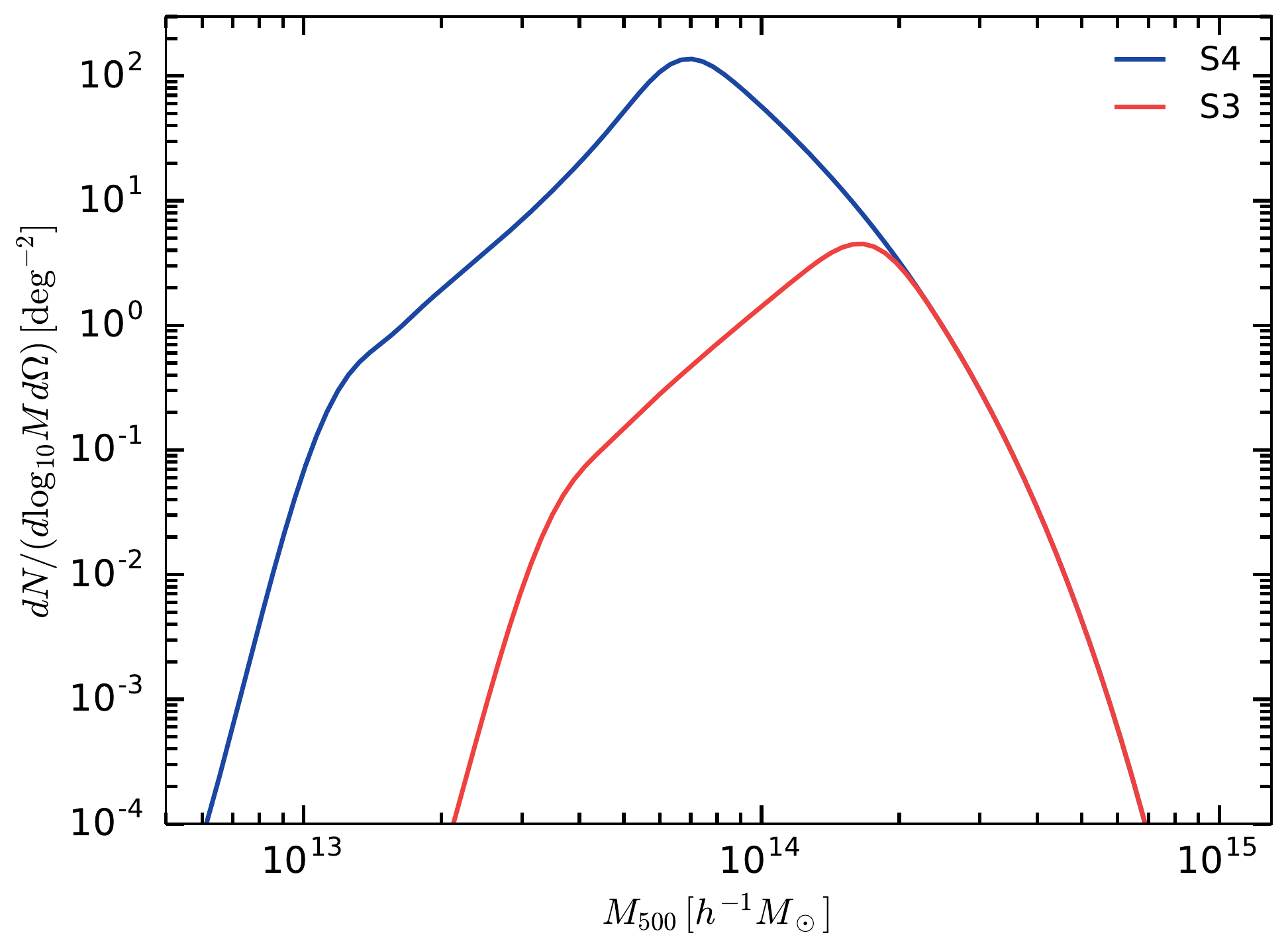}
      \includegraphics[width=0.45\textwidth]{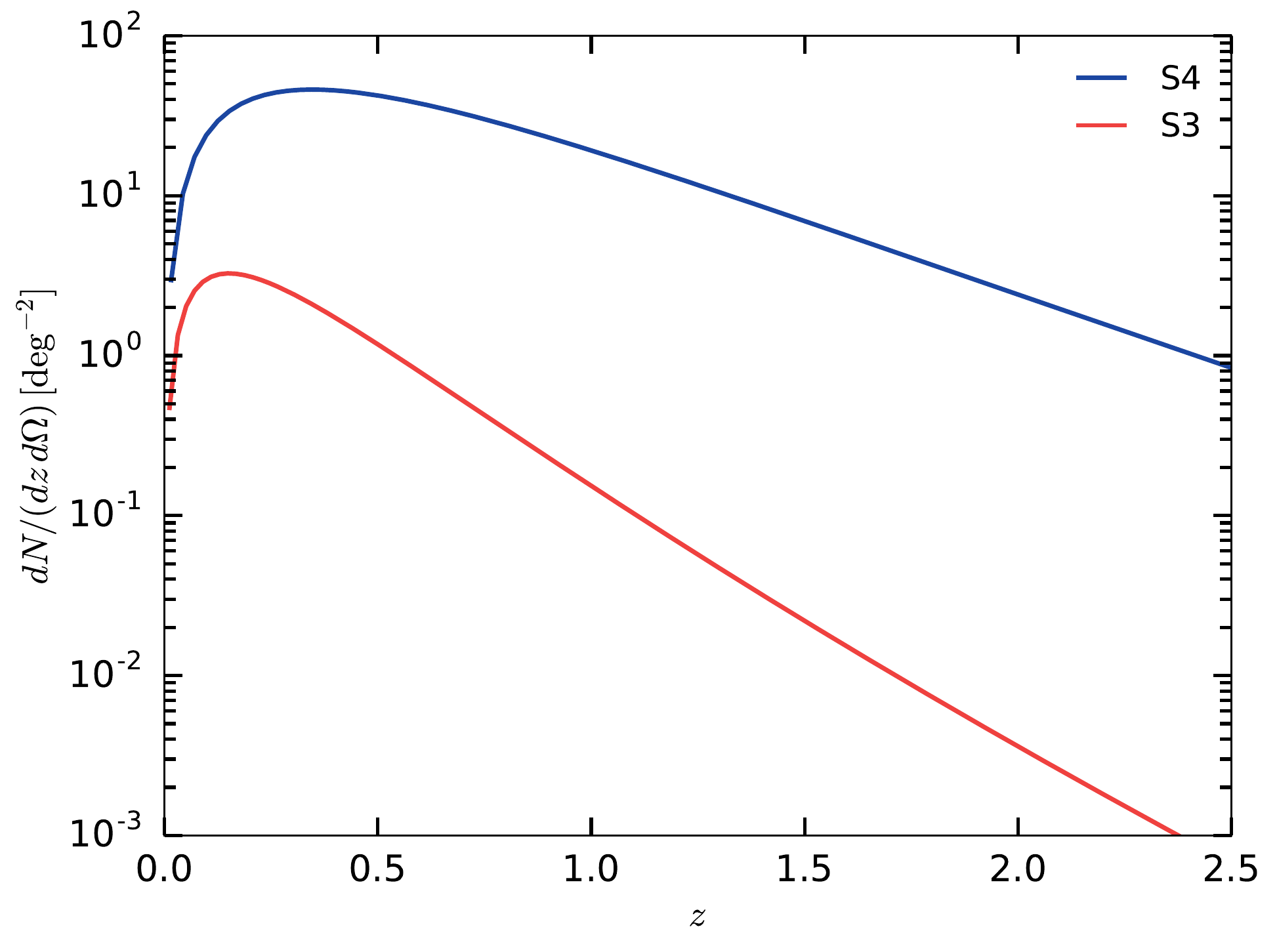}
      \caption{Projected mass (top) and redshift (bottom) distributions of tSZ-selected clusters
               for S3 (red) and S4 (blue).}
      \label{fig:cluster_distributions2}
    \end{figure}

%-------------------------------------------------------------------------------
  \subsection{SZ catalogue properties}
    Using the formalism in Sect.~\ref{ssec:tsz_clusters}, we predicted the expected mass and
    redshift distribution of the tSZ-selected cluster catalogues for each experiment
    (Fig.~\ref{fig:cluster_distributions}). Integrated mass and redshift distributions are
    shown in Fig.~\ref{fig:cluster_distributions2}.

    For both S3 and S4, we assumed a $S/N$ threshold for cluster detection of $q=6$\footnote{Using
    a smaller signal-to-noise threshold would yield a larger catalog, containing fainter and less
    massive objects. However, the kSZ signal-to-noise of those new objects is
    equally reduced, and their contribution to the final constraint on $\alpha$ is negligible.
    We verified this by repeating the analysis for $q=4$}, yielding
    catalogues containing $\sim20,000$ and $\sim600,000$ sources respectively (in agreement with
    e.g. \cite{2015arXiv151002809H}). For S3, the bulk of the sample lies in the mass range
    $\log_{10} M_{500}/(h^{-1} M_\odot) \in(13.7,14.5)$, and at redshifts $z\lesssim0.6$, while
    S4 would be able to extend these ranges to $\log_{10}M_{500}/(h^{-1} M_\odot)\gtrsim13.2$
    and $z\lesssim1.5$.
    We validated this calculation by running our forecast pipeline with the specifications of
    the Planck survey \cite{2014A&A...571A...1P}, obtaining a catalogue with properties (mass
    and redshift distributions) similar to the one presented in \cite{2015A&A...581A..14P}.

    Note that the average cluster size projected on the sky for S4 (given the size of the
    instrumental beam) is $\sim4'$, while the expected number density of clusters for S4 is
    large ($\sim35\,{\rm deg}^{-2}$). It is straightforward to show that a fraction
    $f_{\rm blend}\approx 45\%$ of such a sample would overlap with other clusters on the sky
    ($f_{\rm blend} \approx 5\%$ for S3). Although the problem of cluster blending
    could in principle be overcome by using information about the cluster profiles, we have
    taken a conservative approach here and simply multiplied the number density of SZ sources
    by the expected fraction of non-overlapping clusters, $1 - f_{\rm blend}$, essentially
    discarding the blended objects.

%-------------------------------------------------------------------------------
  \subsection{Comparison of kSZ extraction methods}\label{ssec:ksz_methods}
    
    We now compare the performance of the three different kSZ extraction methods described in
    Sect.~\ref{ssec:method_ksz}: matched filtering (MF), constrained realizations
    (CR), and aperture photometry (AP).
    
    \begin{figure}[t]
      \centering
      \includegraphics[width=0.5\textwidth]{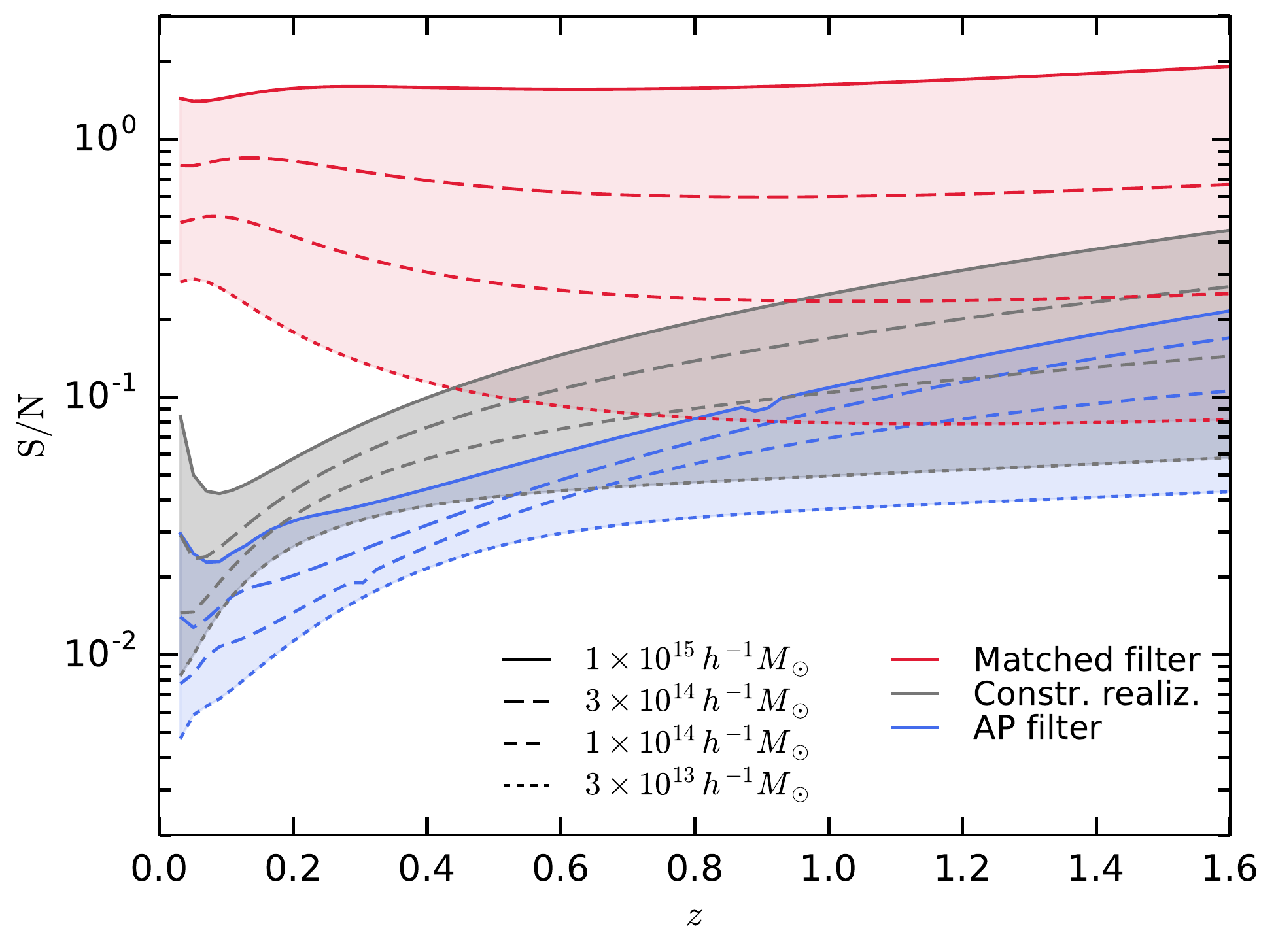}
      \caption{Signal-to-noise ratios for the 3 different kSZ measurement methods, as a
               function of cluster redshift and mass.}
      \label{fig:sn_methods}
    \end{figure}

    Fig.~\ref{fig:sn_methods} shows the signal-to-noise ratio (SNR) of the kSZ amplitude measured
    for a set of characteristic cluster masses and redshifts, assuming a radial velocity
    $v_r=\,300\,{\rm km}/{\rm s}$. For a fixed mass, clusters subtend a larger angle on the sky with
    decreasing redshift, and the performance of the AP and CR methods is therefore significantly
    degraded at low $z$, as larger-scale CMB modes (which have larger variance) enter the filter
    region. This behavior is not reproduced by the MF method, as knowledge of the SZ profile shape
    allows the cluster to be efficiently distinguished from CMB anisotropies, regardless of its
    increased variance. In fact, the MF method sees a slight increase in SNR at low redshift for
    low mass clusters, as the relatively weak signal can be added up coherently over a larger
    number of pixels.

    The CR method shows a definite improvement over AP for all masses and redshifts, typically
    gaining a factor of $\sim\!2$ in SNR. While this is a factor of between $3-20$ worse than the
    MF method, it is nevertheless a significant improvement for a model-independent method,
    especially considering the increased precision on the cosmological parameter measurement
    (see Sect.~\ref{ssec:constraints}).

    \paragraph*{Profile uncertainty:} While MF has by far the best performance in our simulations,
    its efficacy relies heavily on the accuracy of the assumed cluster profile. Given the current
    large uncertainty on the mean profile shape (e.g. see \citep{Schaan:2015uaa}), and the typical
    scatter in the profile from
    cluster to cluster, it is important to fold profile uncertainties into the errors on the
    recovered velocity. This is often achieved by repeating the analysis over a grid of profile
    parameter values for each cluster, although this rapidly becomes impractical as the number of
    parameters grows. Alternatively, a Monte Carlo parameter sampling approach can be taken, as
    discussed in \citep{2015ApJS..219...10B}.

    \begin{figure}[t]
    \centering
    \includegraphics[width=0.485\textwidth]{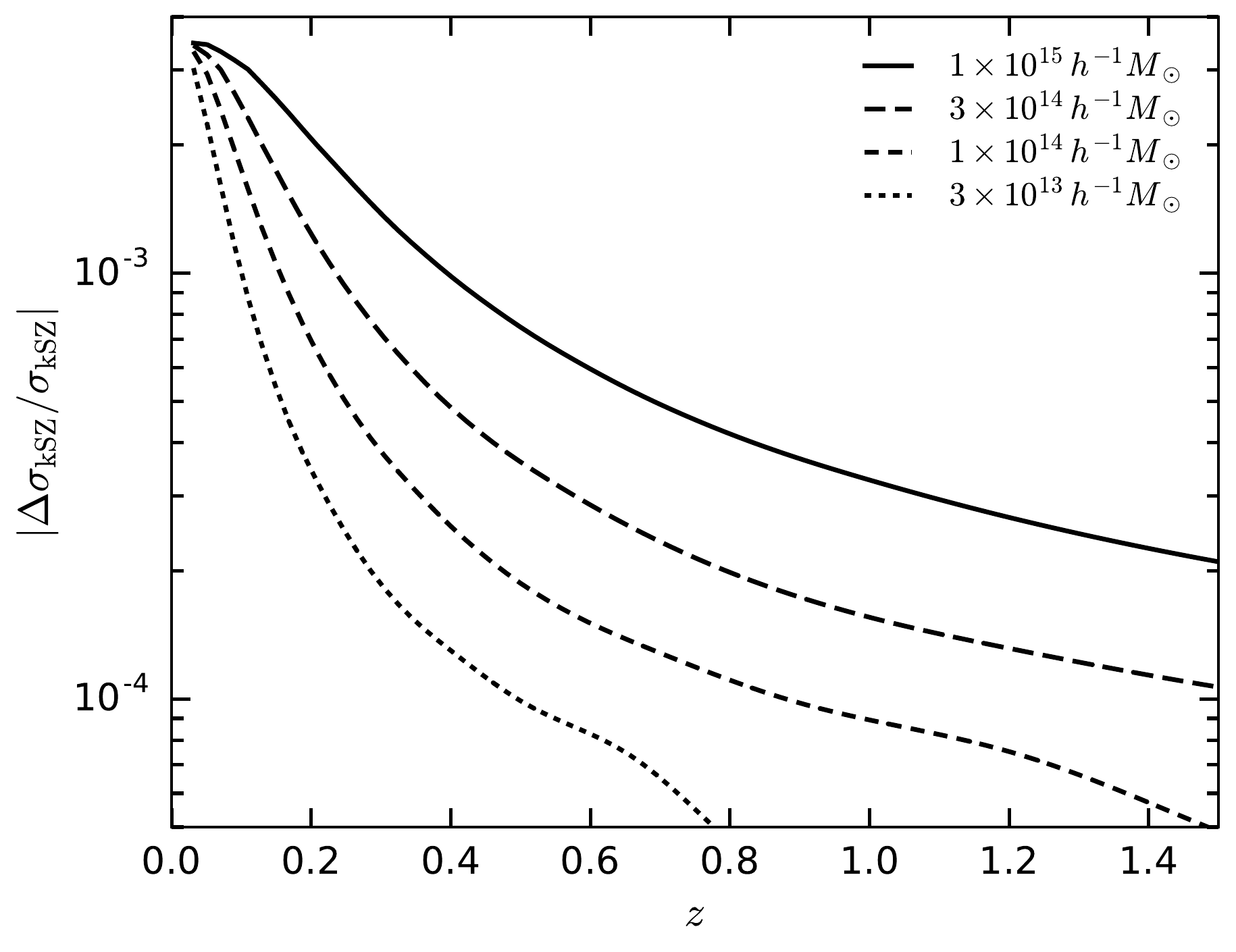}
    \caption{Improvement in the kSZ measurement error for S3 after including polarization data.
             The marginal improvement for large clusters is due to the non-zero
             correlation between $T$ and $E$, and is negligible overall.}
    \label{fig:sn_pol}
    \end{figure}
    
    The profile parameters are often poorly constrained however, and can suffer from strong
    degeneracies. We checked that this is likely to be the case by calculating Fisher
    matrices for the parameters of the GNFW profile (see Appendix~\ref{app:sz_models} for
    definitions), as constrained by the combined tSZ and kSZ profiles. There are several
    near-degeneracies in the Fisher matrix for S3 and S4, almost independent of redshift.
    After inversion, we find $M_{500}$, $c_{500}$ (the concentration parameter), and
    $\gamma$ (the outer slope) to be most strongly correlated with the cluster velocity,
    with correlation coefficients ranging from $|r| \simeq 0.6 - 0.9$ for a
    $10^{15} h^{-1} M_\odot$ cluster over a range of redshifts. Other parameter
    degeneracies make the matrix near-singular, however. Auxiliary information on the
    cluster shape (e.g. from X-ray observations or galaxy surveys) must therefore be
    added to break degeneracies in a real analysis. This typically relies on the use of
    scaling relations and simulations, the accuracy of which must also be folded into the
    uncertainty -- in lieu of a generic procedure for doing this, we leave a quantitative
    analysis of profile shape uncertainties to future work.

    \paragraph*{Polarization:} Both S3 and S4 are sensitive to polarization as well as total
    intensity. The tSZ and kSZ signals are expected to be almost completely unpolarized, while
    the CMB is not. Furthermore, the $T$ and $E$ CMB anisotropies are correlated, suggesting
    a possible way to improve the CMB reconstruction by including polarization information in
    the methods described in Sect.~\ref{ssec:method_ksz}. As an example, we take the matched
    filter (MF) method and extend the profile matrix ${\rm {\bf U}}$ in
    Eq.~\ref{eq:matched_covariance} with polarized channels in which the SZ profiles are set
    to zero. The variance of the kSZ amplitude is then computed as in
    Eq.~\ref{eq:matched_covariance}, where the noise covariance now contains all auto- and
    cross-correlations between the temperature and polarization channels.

    Fig.~\ref{fig:sn_pol} shows the fractional change in the error on the kSZ measurement due
    to the inclusion of polarization information for S3. The improvement is negligible for all
    relevant cluster masses and redshifts. This result is disappointing but understandable:
    while the non-zero $T-E$ correlation does make it possible to better predict properties
    of the temperature field from the measured polarization field, the correlation is relatively
    small ($\sim10\%$), and basically negligible for noise-dominated scales (corresponding to
    most of the cluster sample).
    
    \begin{figure}[t]
      \centering
      \includegraphics[width=0.5\textwidth]{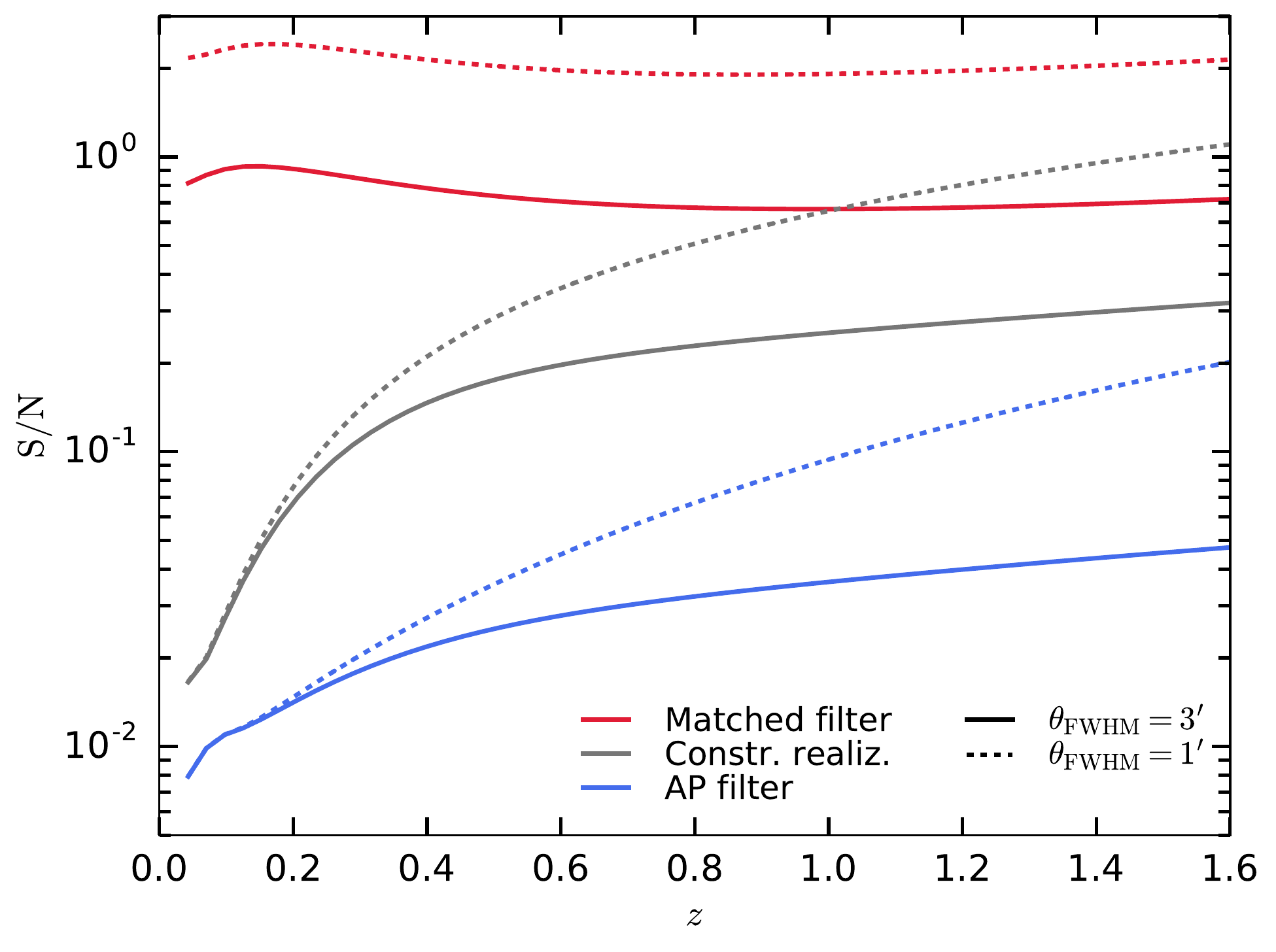}
      \caption{kSZ signal-to-noise ratio for a cluster with mass
      $M_{500}=3\times10^{14}\,h^{-1}M_\odot$ and radial velocity
      $v_r=300\,{\rm km}/{\rm s}$ for S4 with two different beam FWHM: 3 arcminutes
      (solid lines) and 1 arcminute (dotted lines). The uncertainties for the matched filter
      approach improve by a factor $\sim3$ at all redshifts, while the improvement for
      constrained realizations and AP filtering improves gradually at higher redshifts, due
      to the smaller effective angle subtended by the cluster.}
      \label{fig:sn_s4beam}
    \end{figure}

    \paragraph*{S4 specification:}  Finally, as the specification of S4 is uncertain, it is
    worth exploring the posible benefit of different design strategies. For SZ cluster science,
    a narrower instrumental beam would allow the detection and characterization of less-massive
    and more-distant sources. Fig.~\ref{fig:sn_s4beam} shows the dependence of the kSZ
    uncertainty on the S4 beam FWHM for a $3\times10^{14}\,h^{-1}M_\odot$ cluster with
    $c\beta_r=300\,{\rm km}/{\rm s}$. Reducing the beam FWHM for S4 by a factor of $\sim3$
    (i.e. from 3 arcmin to 1 arcmin) would improve the kSZ uncertainties by a similar
    factor at all redshifts for matched filters, and the uncertainty for the cluster-blind
    methods would gradually
    improve to a similar degree towards higher redshifts, where the smaller projected cluster
    size would benefit greatly from a reduced beam size. It is worth noting that, since the tSZ
    uncertainties would be similarly reduced, the effect on the performance of the method described
    here is twofold: first, it would increase the number of tSZ-detected clusters, and second,
    the kSZ uncertainties for those clusters would be reduced.

\begin{figure}[t]
      \centering
      \includegraphics[width=0.5\textwidth]{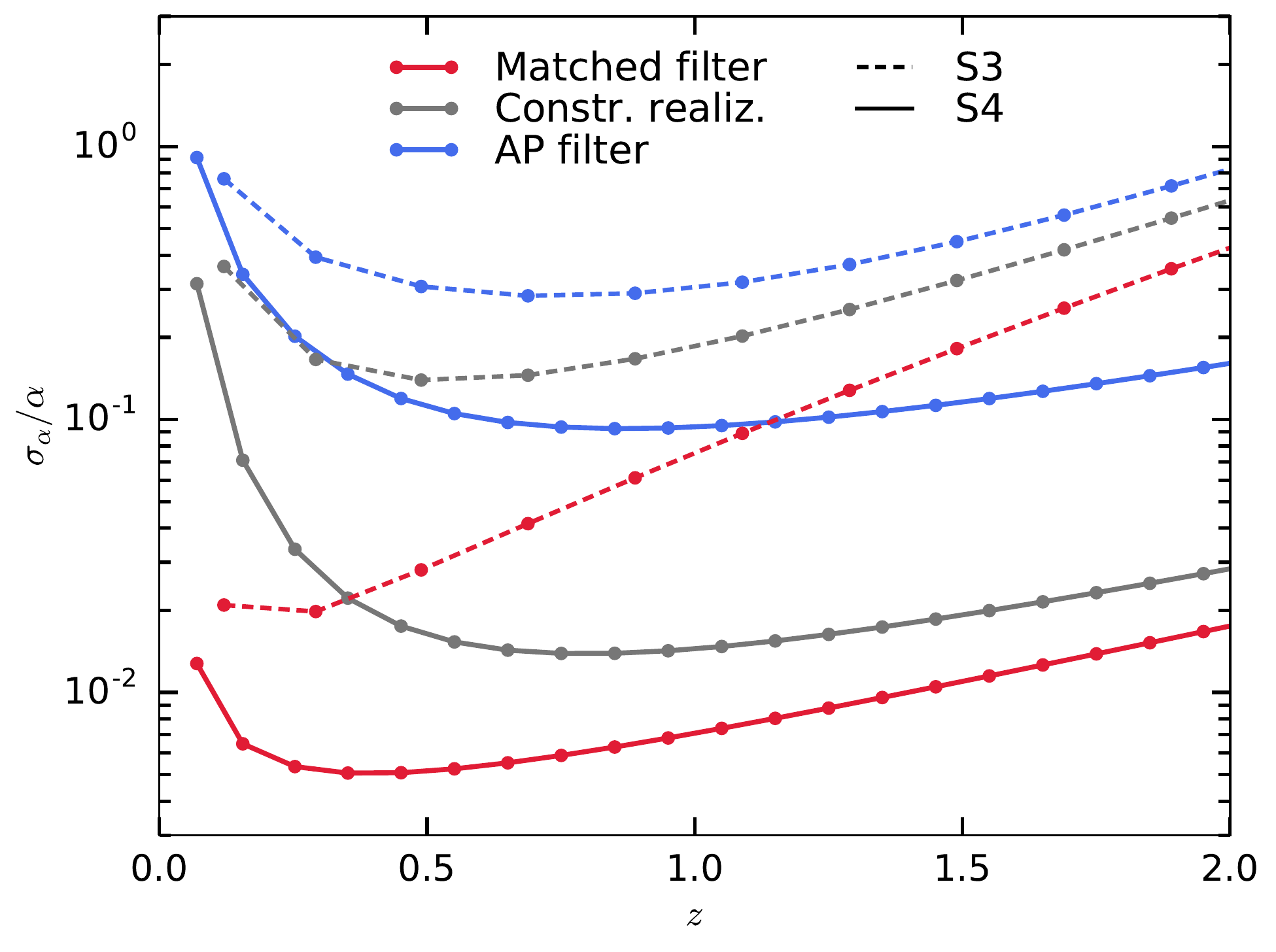}
      \caption{Forecast constraints on $\alpha\sim f\,H$ for S3 (dashed lines) and S4 (solid),
               for the three different methods: matched filtering (MF; red), constrained
               realizations (CR; gray), and aperture photometry (AP; blue). The redshift
               bin widths are $\Delta z=0.2, 0.1$ for S3 and S4 respectively, and we assume
               full redshift and area overlap with a spectroscopic survey.}
      \label{fig:fgh_constraints}
    \end{figure}

%-------------------------------------------------------------------------------
  \subsection{Expansion/growth rate constraints}\label{ssec:constraints}

    We can now combine all of the information from the preceding sections to estimate the
    uncertainty on $\alpha\sim f H$ (Eq.~\ref{eq:errors_alpha}). Fig.~\ref{fig:fgh_constraints}
    shows the forecast relative errors on $\alpha$ for the three kSZ extraction methods for
    both S3 and S4, using redshift bin widths $\Delta z=0.2$ and $0.1$ respectively, and
    assuming full overlap with a spectroscopic galaxy survey.

    As expected, the matched filter (MF) method performs best, with S4 providing extremely
    competitive sub-1\% measurements of $f H$ out to $z=1.5$. The blind CR method is only a
    factor of $2-3$ worse above $z \approx 0.5$, which is also promising, while the AP method
    is a full order of magnitude down, mustering only $\sim 10\%$ constraints for S4. The
    story for S3 is more one-sided, with the MF technique achieving $\sim {\rm few}\, \%$
    constraints out to $z \simeq 0.8$, while the CR and AP methods reach only $\sim {\rm few}
    \times 10\%$ at best. The MF method performs especially well
    at low redshift, where clusters can be well-resolved (especially by the high-resolution S3),
    allowing the shape information assumed by the filter to have the fullest effect. The difference
    is less pronounced at high $z$, so using blind methods here may be preferable due to their
    conservatism.

%-------------------------------------------------------------------------------
  \subsection{Dependence on galaxy survey overlap}\label{ssec:galsurv}
  
\begin{figure}[t]
      \centering
      \includegraphics[width=0.5\textwidth]{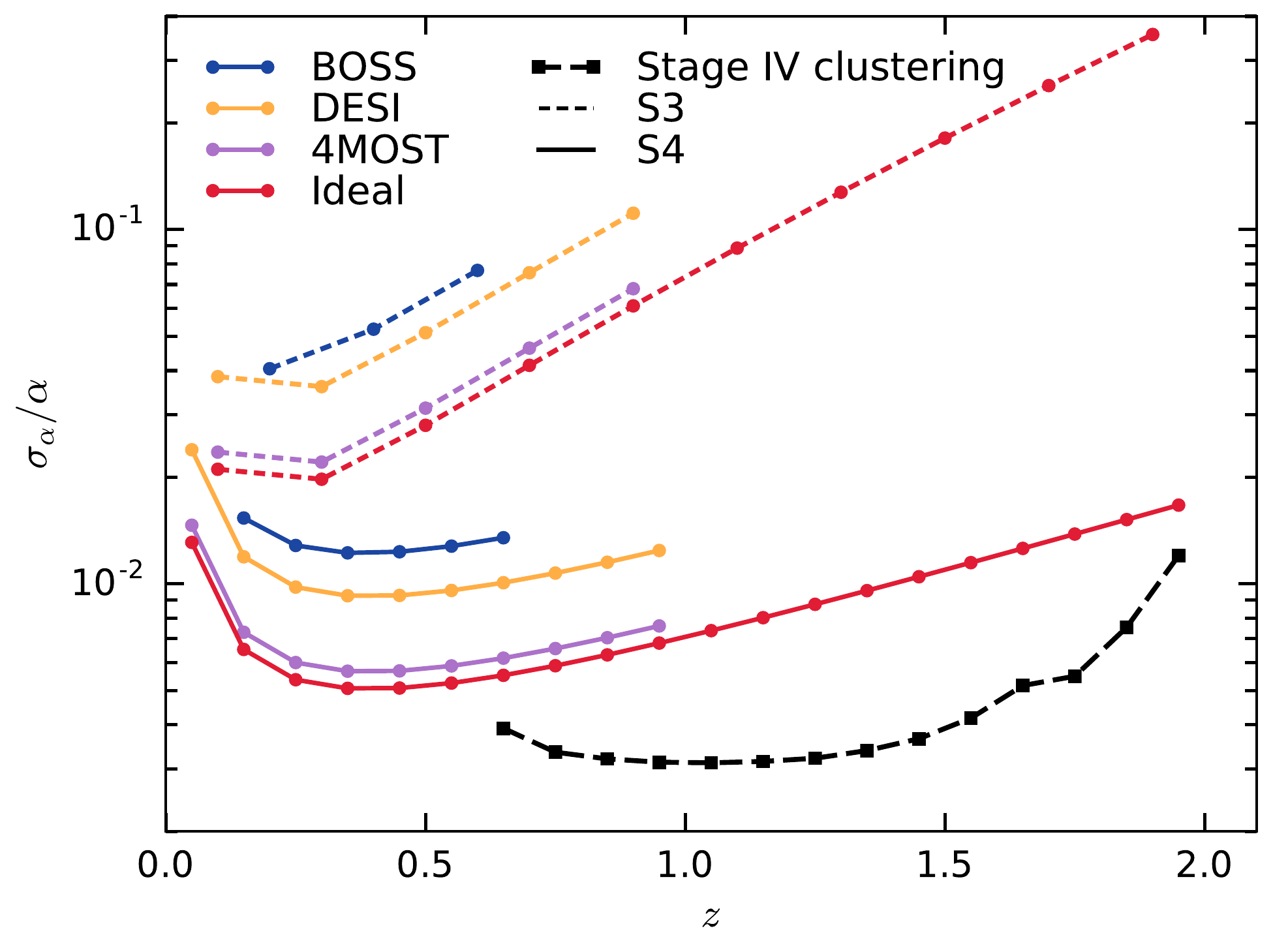}
      \caption{Forecast constraints on $\alpha\sim f\,H$ for S3 and S4 (using the matched filter
      method), when three different galaxy surveys are used to provide the reconstructed velocity
      field. Results for an ideal (perfectly overlapping, sample variance-limited) survey are
      shown in red (c.f. Fig.~\ref{fig:fgh_constraints}). Projected constraints on $\alpha$ from
      BAO + RSDs with a Euclid-like spectroscopic galaxy redshift survey are shown in black.}
      \label{fig:fgh_surveys}
      \vspace{-1em}
  \end{figure}

    The constraints on $\alpha$ ultimately depend on the availability of an overlapping
    spectroscopic galaxy redshift survey. To explore the importance of this issue, we
    selected three forthcoming galaxy surveys according to their expected time of completion:
    BOSS, DESI, and 4MOST. The forecast uncertainties for each of them are shown in
    Fig.~\ref{fig:fgh_surveys}, assuming the
    matched filter method for kSZ extraction. When estimating the overlap of these surveys with our
    model CMB experiments, we have assumed that both S3 and S4 will be southern hemisphere facilities.
    For comparison, the figure also includes the constraints for an ``ideal'' experiment, with full
    redshift and area overlap. Optimistic forecasts for $f H$ from a Euclid-like spectroscopic
    galaxy survey, made by combining BAO and RSD Fisher forecasts from \cite{Bull:2015lja},
    are also shown for comparison.

    The most competitive existing spectroscopic survey, in terms of surveyed volume, is SDSS-III's
    Baryon Oscillation Spectroscopic Survey (BOSS) \citep{2015ApJS..219...12A}. The combination of
    its LOWZ and CMASS samples covers most of the redshift range out to $z=0.7$ over
    $\sim10,000\,{\rm deg}^2$ on the sky, with a number density $n_g\sim10^{-4}\,(h^{-1}{\rm Mpc})^{-3}$.
    We assume a $\sim\!50\%$ area overlap ($5,000\,{\rm deg}^2$) between BOSS and our model S3 experiment,
    due to the northern hemisphere location of BOSS. The relatively low number density of sources in
    BOSS could severely affect the uncertainty on the reconstructed velocities, and so we
    conservatively doubled the size of the uncertainty $\varepsilon_{\beta_r}$ in
    Eq.~\ref{eq:error_reconst}. This is in agreement with the results of \cite{Schaan:2015uaa}. Using
    a matched filter approach, an S3 experiment overlapping with BOSS could obtain a $\sim5\%$
    measurement of the combination $f H$ in the range $0.1<z<0.7$, improving by a factor of $\sim3$
    for S4.

    BOSS will be superseded by the Dark Energy Spectroscopic Instrument (DESI)
    \citep{2013arXiv1308.0847L}, which will operate for 4 years starting in 2018. Jointly, its LRG 
    and ELG samples will cover a similar fraction of the sky to BOSS, but now reaching out to
    $z \simeq 1.5$, and with a higher number density. We assume the same $50\%$ overlap with the S3
    and S4 surveys. Note that the number density will likely be too low to yield a reliable velocity
    field reconstruction in the high-$z$ tail, and so we have only considered the redshift range
    $z<1$ here. The larger number density and redshift coverage of DESI yields a small improvement
    in the forecast uncertainties on $\alpha$ compared to BOSS, with errors of $\sim5-10\%$
    achievable with S3, improving by a factor of $\sim4-10$ for S4.

    The 4-metre Multi-Object Spectroscopic Telescope (4MOST) \citep{2014SPIE.9147E..0MD} will carry
    out a similar spectroscopic survey to DESI in terms of area, depth, and number density, but
    from the southern hemisphere. Although 4MOST will not start operations until 2021, its near-total
    overlap with the survey areas of southern hemisphere CMB experiments such as AdvACT makes it
    ideal for this kind of analysis. We assumed an $80\%$ area overlap ($\sim\!14,000\,{\rm deg}^2$)
    with S3 and S4, and a redshift overlap for all $z<1$. The main improvement over DESI lies in the
    larger area, which translates into a factor of $\sim\!2$ lower uncertainties on $f H$. This
    signal-to-noise level would make these measurements competitive with forecast RSD and BAO
    uncertainties for Stage IV galaxy surveys.

    Finally we note that in the next decade, radio facilities such as the Square Kilometre Array
    (SKA) \citep{2015arXiv150104076M} will carry out spectroscopic galaxy surveys using the 21cm
    radio line. Since any survey carried out by the SKA and its pathfinders would have almost
    complete overlap with both S3 and S4, it is worth exploring the constraints achievable by
    these surveys. Phase 1 of the SKA would be able to produce a $5,000\,{\rm deg}^2$ survey with
    significant number densities out to $z \approx 0.4$ \citep{2015MNRAS.450.2251Y}. The constraints
    from this experiment would therefore be similar to those of DESI for this reduced redshift
    range. The survey would be extended during Phase 2 of SKA to cover $\sim30,000\,{\rm deg}^2$
    out to to $z \simeq 1.3$. The results for such a survey would be similar to those forecast for
    4MOST.

%-------------------------------------------------------------------------------
\section{Discussion}\label{sec:discussion}
  We have studied the potential of measuring the growth rate using a combination of a
  reconstructed velocity field from a galaxy redshift survey and CMB observations of the kSZ
  effect. The performance of this approach depends on the uncertainties with which three
  quantities can be measured: the kSZ flux of each cluster, the cluster velocity reconstructed 
  from the galaxy density field, and
  the cluster optical depth. Of these, we have found the kSZ measurement error to be the
  dominant source of uncertainty for most redshifts and masses, and so we have delved
  deeper in the details of kSZ extraction.
    
  To this end, we have discussed and compared three different methods to measure the kSZ
  with varying degrees of conservatism:
  matched filters (MF), which assume knowledge of both the CMB anisotropies and the mean
  cluster profiles; constrained realizations (CR), which only assume a model of the
  CMB statistics; and angular photometry filters (AP), which separate the primary CMB and kSZ
  components using only qualitative assumptions about their scale dependence.
  
  We have shown
  that these assumptions have a critical effect on the 
  resulting kSZ uncertainties: while AP filters cannot be used to obtain interesting
  constraints on $\alpha\sim f\,H$, constrained realizations could reduce the kSZ
  uncertainties significantly,
  yielding percent-level errors on this quantity assuming a perfectly overlapping
  galaxy redshift survey. Knowledge about the cluster profiles is necessary to
  reduce the uncertainties further, especially at low redshifts where clusters subtend
  larger solid angles. In this case, we have shown that with matched filtering, 
  it would be possible to obtain kSZ
  errors small enough to make this method competitive with RSD-based measurements of the
  growth rate, which should yield sub-percent uncertainties with Stage IV galaxy surveys.
  We have further shown how this method can be extended to make use of polarization data,
  although the level of improvement caused by the $T-E$ correlation in this case is
  negligible. 
  
  It is worth noting that the CR method we propose in this work, based on subtracting our
  best guess of the CMB anisotropies, can significantly improve the S/N of kSZ 
  measurements compared with the commonly-used AP filter, but without 
  requiring strong assumptions to be made about the shape of the cluster SZ profile (as is the 
  case with MF). Although the CR method does require the CMB power spectrum to be specified,
  we are now at a point where it is known with sufficient precision to make this method
  practical.

  The methods presented here build on a number of assumptions. While these should mostly be quite
  reasonable, it is worth bearing in mind the following caveats that will affect any future
  analysis with real data:
  \begin{itemize}
    \item The effectiveness of the matched filter technique depends strongly on the uncertainty in
          the assumed kSZ profile. Marginalizing over profile parameters (e.g. using MCMC sampling
          techniques) is difficult due to the strong degeneracies between most parameters, so
          high-quality external data (e.g. from X-ray and optical observations) is needed to
          better constrain the profile shapes.
    \item The large number densities of clusters that will be detectable means that blending
          (overlapping clusters on the sky) will be an important problem -- several tens of
          percent of clusters will be blended for S4. We have assumed that blended clusters
          can be identified and discarded.
    \item We have ignored several potential biases and uncertainties in the velocity field
          reconstruction procedure, due to effects such as shot noise, RSDs, and non-linear
          and scale-dependent bias. Although the uncertainty on the kSZ measurements should
          dominate the overall error bar, the impact of these effects should be studied in 
          depth. This is the subject of ongoing work.
    \item We have ignored biases and contamination due to imperfect foreground subtraction. Some
          foregrounds (e.g. radio point sources, or the cosmic infrared background) are
          correlated with cluster positions, and so may not average down. It should,
          however, be possible to clean these foregrounds using their different frequency
          spectra.
    \item We have ignored sources of the kSZ effect that are not associated with clusters, such
          as the Ostriker-Vishniac effect from the diffuse IGM, and patchy kSZ from the epoch of
          reionization.
    \item We have only quantified the statistical uncertainties in the three
          observables $(a_{\rm kSZ},\beta_r,\tau_{500})$, neglecting any systematic errors
          in their measurement. The power of this method relies on averaging over many
          low-significance, single-cluster measurements of $\alpha$ by using large numbers of
          clusters. Systematic uncertainties do not average down however, and so, for 
          a sufficiently large number of clusters, the method will eventually be dominated
          by them. This is particularly relevant for one of the key assumptions we
          have made: the existence of a well-calibrated $Y_{500}-\tau_{500}$ relationship,
          needed to break the $\tau-\beta_r$ degeneracy. Due to our imprecise current knowledge
          of cluster physics, systematic deviations can be expected at first, which will
          need to be correctly quantified.
  \end{itemize}

  An important aspect of this method is its different dependence on cosmic variance with respect
  to traditional clustering-based measurements of the growth rate. The statistics of a single
  realization of the density field can only be determined up to an accuracy defined by the number
  of modes accessible in a given survey region. This sample variance limit is easily reached
  by galaxy surveys, given a sufficiently high number density of sources. The
  performance of the method discussed here depends on different factors, however: the measurement
  errors $\varepsilon_i$, and the total number of SZ clusters for which this
  measurement can be carried out. The latter is, in turn, determined by the shape and redshift
  dependence of the mass function and the total surveyed volume. Both sources of uncertainty can 
  (in principle) be reduced without limit, by improving experimental parameters such as the noise 
  sensitivity and angular resolution. This reduces the measurement uncertainties, and extends 
  the mass range of the resulting cluster sample to smaller masses (although, for a fixed lower
  mass bound, the method will be limited by the number of halos present in the surveyed patch,
  which is a different manifestation of the cosmic variance problem). Note that this very fact also
  distinguishes this method from other procedures proposed in the literature to measure the kSZ
  effect, such as the pairwise kSZ signal \cite{1999ApJ...515L...1F} or the projected-field
  probe of \cite{2016arXiv160301608H}.
  
  This leads to an almost complete immunity to cosmic variance, which can be interpreted as 
  follows: the parameter $\alpha \sim f(z) H(z)$ is measured from the combination of two different 
  proxies for the {\it same} velocity field, and so the stochastic velocity terms essentially
  cancel out. This leaves behind a deterministic term that can be measured to arbitrary precision,
  limited only by the aforementioned sources of noise that go into the $\alpha$ estimator, and 
  not by mode counting. A similar effect arises when two differently-biased tracers of the 
  density field are combined to measure RSDs \citep{2009JCAP...10..007M}.

  Due to their tSZ selection functions, and the choice of overlapping galaxy redshift surveys, the 
  growth constraints from S3 and S4 will be mostly restricted to $z \lesssim 1$. This is exactly 
  the regime in which the growth rate has the most to tell us, though -- $f$ deviates increasingly 
  from unity at later times, when dark energy begins to dominate the expansion history. Precision 
  measurements of both growth and expansion at these redshifts are vital to attempts to 
  characterize dark energy and possible modifications of GR. The combination of the two, $\alpha$, 
  constrained by the method described here, is highly complementary to other combinations 
  measured by probes such as BAO and RSDs. By probing the velocity field in a very different 
  (and more direct) way, this method also provides a useful consistency check on RSDs, which use 
  the 2D shape of the clustering pattern, and require a number of modelling assumptions.
  While a successful application of this method will need an excellent
  calibration of systematic uncertainties (especially those related to cluster gas physics),
  we have shown that combined kSZ and galaxy redshift survey analyses promise to become an
  important window into gravitational physics on large scales in the near future.

\section*{Acknowledgments}
  We thank the Nicholas Battaglia, Jo Dunkley, Simone Ferraro, Sigurd N{\ae}ss, and Emmanuel Schaan
  for useful comments and discussion. DA is supported by the Beecroft Trust and ERC grant 259505.
  TL is supported by ERC grant 267117 (DARK) hosted by Universite Pierre et Marie Curie- Paris 6.
  PB's research was supported by an appointment to the NASA Postdoctoral Program at the Jet
  Propulsion Laboratory, California Institute of Technology, administered by Universities Space
  Research Association under contract with NASA. PGF acknowledges support from STFC, the Beecroft
  Trust and the Higgs Centre in Edinburgh.
 
\bibliography{paper}

\appendix
\section{SZ profiles and amplitudes}\label{app:sz_models}
  
\begin{figure}[b]
    \centering
    \hspace{-2em}\includegraphics[width=0.5\textwidth]{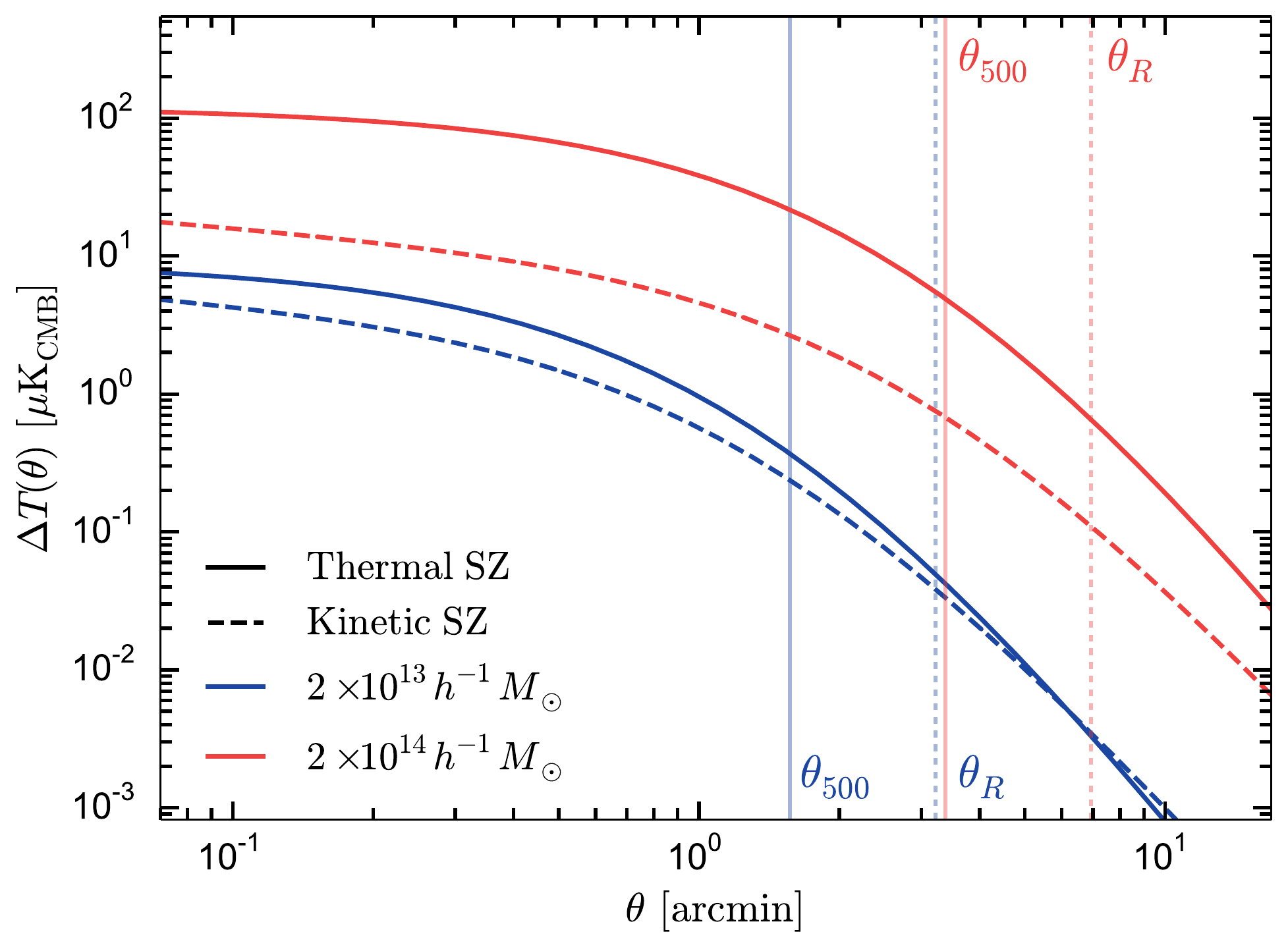}
    \caption{The tSZ (solid lines) and kSZ (dashed) profiles for clusters with
             halo masses $2\times10^{13} h^{-1} M_\odot$ (blue) and
             $2\times10^{14} h^{-1} M_\odot$ (red), both at $z=0.3$ with a radial
             velocity $v_r=-400\,{\rm km}/{\rm s}$. The masses are chosen to be broadly
             representative of the S4 and S3 samples respectively
             (see Fig.~\ref{fig:cluster_distributions2}). The vertical lines
             show the characteristic angular scale $\theta_{500}$ (solid), and the disc
             radius $\theta_R$ (dotted) for the AP filter for a Stage 3 experiment
             ($1.4$ arcmin beam), defined in Section~\ref{sssec:semi_blind}.}
    \label{fig:profiles}
  \end{figure}

  In this appendix we describe the models that were used to estimate the amplitude and
  projected cluster profiles for the tSZ and kSZ components throughout this paper.
  Fig.~\ref{fig:profiles} shows examples of the profiles for two different cluster
  masses at $z = 0.3$.
  
  \subsection{Thermal SZ profile}
    The tSZ contribution to the CMB anisotropies is given by Eq.~\ref{eq:tsz}, with
    \begin{equation}
      f_{\rm tSZ}(\nu)\equiv\frac{q(e^q+1)}{e^q-1}-4,
      \hspace{12pt}q\equiv\frac{h\nu}{k_BT_{\rm CMB}}.
    \end{equation}
    To construct our model, we assume that the tSZ pressure profile is well described
    by the GNFW/Arnaud profile \cite{2010A&A...517A..92A}, i.e.
    \begin{align}\label{eq:arnaud}
     &\frac{\sigma_T k_{\rm B}}{m_ec^2} n_e(r) T_e(r) = L_0^{-1}\,p_p(r/R_{500}) \\
     &p_p(x)=\left[(x\,c_{500})^\gamma\,
     [1+(x\,c_{500})^\alpha]^{(\beta-\gamma)/\alpha}\right]^{-1},
    \end{align}
    where $L_0$ is a constant prefactor (with units of length), $p_p(x)$ is the
    dimensionless pressure profile, and the profile parameters are the best-fit
    values from \citep{2010A&A...517A..92A}: $c_{500}=1.156$, $\alpha=1.062$,
    $\beta=5.4807$, $\gamma=0.3292$. Now, define the tSZ flux $Y_{500}$ as
    \begin{equation}
      Y_{500}=\frac{4\pi}{d_A^2}
      \int_0^{R_{500}}dr\,r^2\,n_e(r)\frac{k_BT_e}{m_ec^2}\sigma_T.
    \end{equation}
    We can then write the tSZ anisotropy as in Eq. \ref{eq:model1}, with
    \begin{align}
      &a_{\rm tSZ}\equiv Y_{500},\hspace{12pt}
      u_{\rm tSZ}(\nu,\theta)=f_{\rm tSZ}(\nu)\,
      \frac{g_{\rm tSZ}(\theta/\theta_{500})}{4\pi\theta_{500}^2}
    \end{align}
    \begin{align}
      &g_{\rm tSZ}(x)\equiv\frac{\int_{-\infty}^\infty dx_z\,p_p\left(\sqrt{x_z^2+x^2}\right)}
      {\int_0^1 dx_r\,x_r^2\, p_p(x_r)},
    \end{align}
    where $x_r$ denotes the radius from the centre of the cluster, and $x_z$ is the distance
    along a line of sight through the cluster (with closest approach to the centre,
    $x_z = 0$, at a radius $x$).
      
  \subsection{Kinetic SZ profile}
    The kSZ profile is determined by the electron density rather than the pressure profile.
    Here we will model $n_e$ by assuming that the cluster is in hydrostatic equilibrium
    \cite{2016MNRAS.455.2936S},
    \begin{equation}\label{eq:ne}
      n_e(r)=\frac{\rho_{\rm gas}}{m_p\mu_e}=-\frac{r^2}{GM(<r)m_p\mu_e}\frac{dP_e}{dr},
    \end{equation}
    where $\rho_{\rm gas}$ is the baryon mass density, $M(<r)$ is the total matter enclosed in a
    sphere of radius $r$, $P_e$ is the electron pressure and $\mu_e$ is the mean molecular
    weight per free electron. We assume a mass profile given by the NFW universal halo profile
    \cite{1996ApJ...462..563N},
    \begin{align}
      M(<r)&=M_{500}p_M(r/R_{500}),\\
      p(x)&=
      \frac{\ln(1+c_{500}x)-c_{500}x/(1 + c_{500}x)}
      {\ln(1+c_{500})-c_{500}/(1+c_{500})},
    \end{align}
    and the GNFW presure profile $P_e(r)=m_ec^2/(L_0\sigma_T)p_p(r/R_{500})$ as above.
    Evaluating Eq.~\ref{eq:ne}, we obtain
    \begin{equation}
      n_e(r)\equiv\frac{m_ec^2}{GM_{500}m_p\mu_e\sigma_T}\frac{Y_{500}\, p_n(r/R_{500})}
      {4\pi\theta_{500}^2\int_0^1dx\,x^2\,p_p(x)},\nonumber
    \end{equation}
    where we have defined the dimensionless number density profile
    $p_n(x)\equiv-x^2p_p'(x)/p_M(x)$.

    Now, define $\tau_{500}$, the quantity analogous to $Y_{500}$, as
    \begin{align}
      \tau_{500}&\equiv\frac{4\pi}{d_A^2(z)}\int_0^{R_{500}}dr\,r^2\,n_e(r)
                 \sigma_T\\
                &=\frac{m_ec^2\,Y_{500}\,R_{500}}{GM_{500}m_p\mu_e}
                  \frac{\int_0^1dx\,x^2\,p_n(x)}{\int_0^1dx\,x^2\,p_p(x)}\\
                =&99.8 \,\frac{\int_0^1dx\,x^2\,p_n(x)}{\int_0^1dx\,x^2\,p_p(x)}
                  \left[\frac{R_{500}}{1\,{\rm Mpc}/h}\right]
                  \left[\frac{Y_{500}}{{\rm srad}^2}\right]
                  \left[\frac{M_{500}}{10^{14}\,M_\odot/h}\right]^{-1}.\nonumber
    \end{align}
    The kSZ anisotropy can finally be written as in Eq.~\ref{eq:model1},
    \begin{align}
      &a_{\rm kSZ}\equiv -\beta_r\tau_{500},\hspace{12pt}
      u_{\rm kSZ}(\nu,\theta)=
      \frac{g_{\rm kSZ}(\theta/\theta_{500})}{4\pi\theta_{500}^2},
      \\
      &g_{\rm kSZ}(x)\equiv\frac{\int_{-\infty}^\infty dx_z\,p_n\left(\sqrt{x_z^2+x^2}\right)}
      {\int_0^1 dx_r\,x_r^2\, p_n(x_r)}.
    \end{align}
     \clearpage

\end{document}